\PassOptionsToPackage{hyphens}{url}

\documentclass[a4paper,fleqn]{cas-sc}

\usepackage[table]{xcolor}
\definecolor{cite_color}{rgb}{0.0, 0.58, 0.71}
\definecolor{db}{rgb}{0.0, 0.2, 0.7}
\definecolor{headerBlue}{RGB}{51,122,183}
\definecolor{lightGray}{RGB}{245,245,245}
\usepackage{natbib}
\setcitestyle{authoryear}
\usepackage[T1]{fontenc}
\usepackage[nomath]{lmodern}
\usepackage{amsmath,amsthm,amsfonts,amssymb}
\usepackage{mathtools}
\numberwithin{equation}{section}
\usepackage{booktabs}
\usepackage{array}
\usepackage{graphicx}
\usepackage{subcaption}
\usepackage{longtable}
\usepackage{threeparttable}
\usepackage{float}
\usepackage[section]{placeins}
\usepackage{wrapfig}
\usepackage{enumitem}
\usepackage{tabularx}
\usepackage{ragged2e}

\usepackage{microtype}

\makeatletter
\@ifpackageloaded{hyperref}{}{%
  \usepackage{hyperref}
}
\makeatother
\hypersetup{
  colorlinks=true,
  citecolor=cite_color,
  linkcolor=cite_color,
  urlcolor=cyan,
  filecolor=magenta,
  pdfauthor={},
  pdftitle={}
}

\usepackage{etoolbox}
\usepackage{xspace}

\newif\ifreviewnotes
\reviewnotesfalse
\newcommand{\authorcheck}[2]{%
  \ifreviewnotes\par\smallskip\noindent
  {\small\color{red!65!black}\textbf{Author check #1.} #2}\par\smallskip\fi
}

\makeatletter
\newcommand{\tsc}[1]{%
  \csdef{#1}{\textsc{\lowercase{#1}}\xspace}
}

\makeatother
\tsc{WGM}\tsc{QE}\tsc{EP}\tsc{PMS}\tsc{BEC}\tsc{DE}

\renewcommand{\figurename}{Fig.}
\makeatletter
\renewcommand*{\fnum@figure}[1]{\figurename~\thefigure.}
\makeatother

\renewcommand{\arraystretch}{1.3}
\begin{document}
\let\WriteBookmarks\relax
\let\printorcid\relax
\setlength{\skip\footins}{24pt}
\setlength{\emergencystretch}{1em}
\renewcommand{\topfraction}{0.9}
\renewcommand{\bottomfraction}{0.8}
\renewcommand{\textfraction}{0.07}
\renewcommand{\floatpagefraction}{0.75}
\setcounter{topnumber}{4}
\setcounter{bottomnumber}{3}
\setcounter{totalnumber}{6}

\shorttitle{ADAS-TO: Data and Takeover Behavior}
\shortauthors{Wang et~al.}

\title [mode = title]{
Driver Takeovers and Longitudinal Responses Analysis from Naturalistic Driving
}

\author{Yuhang Wang}
\author{Hao Zhou\corref{corresponding}}
\ead{haozhou1@usf.edu}

\address{Department of Civil and Environmental Engineering, University of South Florida, Tampa, FL 33620, USA}
\cortext[corresponding]{Corresponding author.}

\begin{abstract}
As Level 2 advanced driver-assistance systems (ADAS) become more widely used, understanding driver--system interaction is increasingly important for traffic safety. However, limited real-world data make these interactions difficult to study. We present ADAS-TO, a naturalistic dataset of 16,446 takeover clips from 364 drivers across 2,585 different drives. It combines road video, vehicle kinematics, driver inputs, and ADAS status. 
To analyze takeover behavior, we retained 6,206 manually reviewed takeover events during continued lane keeping or car following. We excluded events in which the driver intentionally changed lanes, turned, or made a stop beyond the ADAS's capabilities (e.g., at a red light or stop sign) shortly after takeover.
Drivers began intervening before ADAS disengaged in 65.6\% of takeovers involving braking or steering. When the ego vehicle was still closing on a lead vehicle at disengagement, longitudinal safety margins often narrowed further after the transfer to manual control. Among 2,031 such events, 78.5\% had a lower minimum time-to-collision (TTC) in the five seconds after disengagement than in the five seconds before it. For events involving braking, the mean TTC trajectory reached its minimum about 1.2~s after disengagement.
We model the subsequent braking response using a Gazis--Herman--Rothery (GHR) car-following law. Its median five-second spacing root-mean-square error is 1.34~m with observed lead motion and 1.84~m with predicted lead motion. The fitted model expresses braking as a response to looming, with stronger responses at higher speeds and shorter gaps. Applied before takeover, it shows that greater looming than predicted by the human-response reference is associated with disengagement. These findings support continued assessment of longitudinal safety after ADAS disengagement, while the driver's braking response develops.
\end{abstract}


\maketitle
\hypersetup{pdfauthor={Yuhang Wang, Hao Zhou},pdftitle={ADAS-TO: A Naturalistic Takeover Dataset, Longitudinal Safety Analysis, and Human-Response Modeling}}

\section{Introduction}
\label{sec:introduction}

Partial driving automation is increasingly available in production vehicles \citep{spglobal2025_l2_availability,spglobal2025_l2_adoption}.
Level-2 systems control vehicle speed and steering while the driver supervises the road and remains responsible for safe driving \citep{sae_j3016_202104}.
Drivers may take over without an explicit request when the assistance no longer responds as they consider appropriate \citep{nordhoff2024disengagement}.
Understanding these takeovers requires examining what happens before the driver intervenes and how the vehicle responds afterward \citep{zhang2025actual,chen2023study,woide2023ve}.

A driver can begin pressing a pedal or turning the steering wheel while ADAS is still active.
The system may disengage only after this initial intervention, depending on its control and disengagement logic.
The first driver action and the switch from ADAS to manual driving therefore need to be distinguished.
This differs from the usual setup in controlled takeover studies, where an explicit request provides a common starting point for measuring the driver's response.
Simulator experiments allow researchers to vary that request and the surrounding traffic, but their risk conditions and system behavior differ from real-world driving \citep{himmels2024simulatorvalidity}.
Naturalistic resources such as MIT-AVT, L3Pilot, AIDE, and CARIM provide complementary observations \citep{fridman2019advanced,perez2024characterizing,yang2023aide,ki2025carim,cooper2023driverbehavior}.
Public datasets that align the road scene, driver actions, ADAS status, and vehicle response throughout a takeover remain limited.

The driving context also matters.
A driver may disengage ADAS to turn, change lanes, stop at an intersection, or respond to a vehicle ahead \citep{morando2020driver,yang2023takeover,schwager2024analysis}.
These situations produce different braking and steering patterns.
We separate planned maneuvers before examining takeovers during continued lane keeping or car following.
For longitudinal takeovers, we follow the approach to the lead vehicle, the driver's intervention, and the evolution of TTC after ADAS disengages.
The driver's initial action may precede the braking needed to reverse a decline in TTC.
Examining this interval can inform safety assessment and collision-avoidance support during the transfer to manual control.

The behavior after takeover also offers a reference for understanding the preceding approach.
It shows how drivers brake in the situation they have just encountered.
We fit a Gazis--Herman--Rothery (GHR) car-following model to these responses and apply it to the preceding assisted-driving period.
GHR provides a compact description of braking in terms of looming, the closing rate relative to the remaining distance, with additional dependence on speed and spacing.
This interpretation connects the model with evidence that the onset and build-up of human braking depend on the urgency of the approach \citep{markkula2016farewell}.
We then examine whether takeover is more likely when the observed approach produces greater looming than this human-response model predicts.

The paper makes three contributions:




\begin{enumerate}[leftmargin=*,itemsep=0.45em,topsep=0.4em]
\item \textbf{A naturalistic ADAS takeover dataset.}
ADAS-TO contains 16,446 takeover clips from 364 drivers, spanning 2,585 drives and 99 vehicle models. Each clip synchronizes road video, vehicle kinematics, driver inputs, and ADAS status before and after disengagement. For the behavioral analysis, we retained 6,206 manually reviewed takeovers during continued lane keeping or car following. We excluded events followed shortly by a intentional lane change, a turn, or a stop beyond the system's capabilities, such as a red light or stop sign. Stops made while following a lead vehicle remained eligible as car-following events.

\item \textbf{Longitudinal safety margins can narrow further after ADAS disengagement.}
Among 2,031 car-following events with a closing gap at disengagement, 78.5\% had a lower minimum time-to-collision (TTC) in the following five seconds than in the preceding five seconds. For events involving braking, the mean TTC trajectory reached its minimum about 1.2~s after disengagement. We also describe the traffic contexts of safety-critical events and estimate time to conflict resolution from the available follow-up.

\item \textbf{Looming-based analysis of braking and disengagement.}
We fit a Gazis--Herman--Rothery (GHR) car-following model to post-takeover trajectories and evaluate it on held-out drivers. During assisted car following, we use the model to simulate one second of human driving from the observed initial state under the recorded lead-vehicle motion. A larger positive difference between observed and simulated looming is associated with subsequent ADAS disengagement.
\end{enumerate}

Section~\ref{sec:related_work} reviews the relevant literature.
Section~\ref{sec:take_over_dataset} introduces the dataset, and Section~\ref{sec:stats} describes takeover behavior.
Section~\ref{sec:model} presents the response model and its relation to disengagement.
Sections~\ref{sec:discussion} and~\ref{sec:conclusion} discuss the findings and conclude the paper.
In the event-aligned analysis, $t=0$ marks the switch from ADAS engagement to disengagement.

\section{Related Work}
\label{sec:related_work}

Three lines of work inform this study: takeover timing and performance, naturalistic ADAS use, and models of longitudinal driver response.

\subsection{Takeover Timing and Performance}

A large body of takeover research relies on driving simulators, which allow controlled manipulation of takeover-request (TOR) timing, non-driving-related tasks, traffic density, urgency, and human--machine interface design. Simulator studies also examine urgency and visibility in critical car-following takeovers \citep{lin2024how}.
Early studies examined how quickly drivers return to control after an explicit TOR \citep{gold2013take}, and later work showed that takeover time alone does not characterize takeover safety, because visual--cognitive load and situation understanding also shape takeover quality \citep{zeeb2016take,zeeb2015what}.
Post-takeover measures include steering, braking, lateral acceleration, obstacle clearance, and time-to-collision \citep{happee2017take}, and recent public resources such as TD2D provide multimodal simulator data for distracted L2 takeover analysis \citep{hwang2025dataset}.

Two findings from this literature motivate the timing analysis.
First, takeover time is strongly condition-dependent: the meta-analysis of \cite{zhang2019determinants} across 129 studies quantifies effects of time budget, secondary tasks, TOR modality, and urgency, and \cite{eriksson2017takeover} report noncritical takeover times spanning 1.9--25.7~s with a heavy right tail, arguing that full distributions rather than point summaries should be reported.
Second, reviews of computational takeover modeling conclude that drivers respond to takeovers much as they do to manual-driving emergencies, with evidence-accumulation models the most promising generative account, while noting that most reporting remains means-based \citep{mcdonald2019toward}.
Much of this evidence uses a TOR as the temporal anchor.
Many driver-initiated Level-2 interventions have no such anchor, so we anchor timing on the logged disengagement and report the lead time of the first driver action relative to it.
Faster is also not necessarily safer: post-takeover metrics such as maximum deceleration, TTC, and headway are needed alongside timeliness \citep{cao2021towards,han2026systematic,sekadakis2025systematic}.

\subsection{Naturalistic ADAS Analysis}

On real roads, Wizard-of-Oz and test-track studies investigated responses to TORs, visual attention, early versus late alerts, and automation aftereffects \citep{pipkorn2022s,pipkorn2021automation,pipkorn2023driver,pipkorn2024driver,doi:10.1177/0018720818824002}, but remain small-sample and scripted.
Naturalistic studies provide broader evidence: the MIT-AVT study collected long-term data from production vehicles with partial automation \citep{fridman2019advanced}, and derived analyses examined Autopilot and Super Cruise engagement, disengagement, gaze, and recovery behavior \citep{morando2020driver,morando2021model,morando2021visual,gershon2023driver,yang2023takeover}; \cite{gershon2021driver} characterized thousands of transfer-of-control cycles and their contexts.
Recent field analyses of L2 systems document disengagement rates, triggers, and driver adaptation \citep{nordhoff2024disengagement,cooper2023driverbehavior,perez2024characterizing,mueller2025finding}, and crash-record studies begin to quantify system-level safety effects \citep{cicchino2025convenience}.
ADAS-TO complements these resources with event-centered clips that align road video, automation state, driver input, and lead-vehicle kinematics.

A parallel empirical literature characterizes how production longitudinal automation itself follows traffic.
Experiments have documented delayed responses and string instability in commercial ACC systems: \cite{milanes2014modeling} model production ACC with a first-order response lag; \cite{gunter2021are} calibrate seven production systems from field data and find all string-unstable; the OpenACC campaigns document response times and equilibrium headways across vehicles \citep{makridis2021openacc}; and empirical studies of automated vehicles report longer response lags and larger headways than human drivers \citep{shi2021empirical,ye2023carfollowing}.
These studies characterize ACC in normal following.
Event-centered data can extend this work by identifying the conflicts preceding takeovers and documenting vehicle behavior as they develop.

\subsection{Longitudinal Response Modeling and Conflict Assessment}

Classical car-following models provide simple, interpretable relations between a traffic situation and the driver's longitudinal response.
The linear feedback law of \cite{helly1959simulation}, the safe-speed model of \cite{gipps1981carfollowing}, the optimal-velocity family \citep{bando1995dynamical}, the parsimonious first-order model of \cite{newell2002simplified}, and the Intelligent Driver Model \citep{treiber2000idm} represent several widely used model families.
Equally important is the calibration literature: goodness-of-fit should be evaluated on simulated spacing rather than instantaneous acceleration, parameter identifiability on short trajectories is limited, and estimated parameters vary substantially across drivers \citep{punzo2012can,punzo2015do,kesting2008calibrating,ossen2011heterogeneity}.
We use a four-parameter GHR law to relate post-takeover following behavior to the preceding approach, and evaluate its five-second spacing trajectories in grouped holdout tests.

Models of human braking initiation give the timing analysis its theoretical frame.
Visual looming has long been studied as a cue for time-to-contact judgments \citep{lee1976theory}; naturalistic analyses show that brake initiation in rear-end conflicts is better described by looming-threshold and evidence-accumulation processes than by fixed reaction times \citep{markkula2016farewell,svard2021computational,xue2018using}.
We express the GHR response in terms of looming and use departures from its calibrated response to characterize behavior before takeover. This connects a familiar closing cue with both the intervention and the subsequent car-following response.

We use established surrogate measures to describe the conflicts: TTC originates with \cite{hayward1972near}, time-exposed and time-integrated TTC extend it to exposure \citep{minderhoud2001extended}, DRAC expresses the required avoidance deceleration \citep{cooper1976traffic}, and \cite{guo2010near} examines near-crashes as crash surrogates in naturalistic data.
These measures summarize particular aspects of a conflict and depend on their observation window and signal validity. We therefore report event-aligned trajectories, symmetric-window extrema, and paired-event comparisons, while using manual context review to distinguish rapid closing from other intervention settings.

The present study combines event-centered naturalistic records with a description of the control-transfer process and a response reference calibrated on manual driving. This addresses how a closing conflict develops across disengagement and whether the response observed afterward helps characterize the approach beforehand.
Table~\ref{tab:dataset_comparison} positions ADAS-TO against existing takeover and driver--automation datasets.

\begin{table}[pos=!htbp]
\centering
\footnotesize
\setlength{\tabcolsep}{3.5pt}
\renewcommand{\arraystretch}{1.15}
\caption{Comparison of ADAS-TO with representative takeover and driver--automation datasets.}
\label{tab:dataset_comparison}
\begin{tabularx}{\linewidth}{@{}p{2.5cm}p{2.1cm}p{2.6cm}X@{}}
\toprule
\textbf{Resource} &
\textbf{Setting} &
\textbf{Scale} &
\textbf{Data and takeover relevance} \\
\midrule

Simulator studies &
Simulator &
Small/medium &
Control traces, gaze, workload; controlled TOR scenarios. \\

TD2D &
Simulator &
50 drivers; 500 cases &
Physiology, ocular signals, workload, takeover metrics; public but simulator-based. \\

Road test-track TOR studies &
Real vehicle &
Small samples &
Gaze and vehicle control; scripted and TOR-centered. \\

MIT-AVT &
Naturalistic roads &
122 participants; 511k miles &
Video, CAN, GPS, IMU; large-scale automation-use data, not takeover-centered. \\

L3Pilot Open Data &
Public roads &
European AD pilots &
Processed indicators; not raw event-level takeover clips. \\

AIDE / CARIM &
Driving videos &
Scene datasets &
Video and semantic captions; useful for scene understanding, not takeover analysis. \\

\textbf{ADAS-TO} &
\textbf{Naturalistic roads} &
\textbf{16,446 clips; 364 drivers; 99 vehicle models} &
\textbf{Road video, CAN-derived signals, DAS state, driver intervention, and semantic labels; event-level takeover data.} \\

\bottomrule
\end{tabularx}
\vspace{2pt}
\footnotesize
\emph{Note.} TOR = takeover request; DAS = driving automation system.
\end{table}


\section{ADAS-TO Dataset and Analysis Samples}
\label{sec:take_over_dataset}

ADAS-TO preserves the road scene and vehicle signals around logged ADAS disengagement. This section describes the recording configurations, the event-centered records, and the samples used for the descriptive and modeling analyses.

\subsection{Data Collection}
\label{sec:dataset_constrcution}

ADAS-TO is built from naturalistic driving logs recorded with an openpilot-based stack \citep{openpilot}, combining in-house recordings from the Tampa metropolitan area with community-contributed drives.
The recording stack plays two roles: in openpilot-controlled operation it generates the vehicle-control commands before the takeover; in OEM-native operation it acts as a synchronized logger while the factory ADAS controls the vehicle.
Both operation types contribute takeovers.

\begin{figure}[pos=!htbp]
    \centering
    \includegraphics[width=\linewidth]{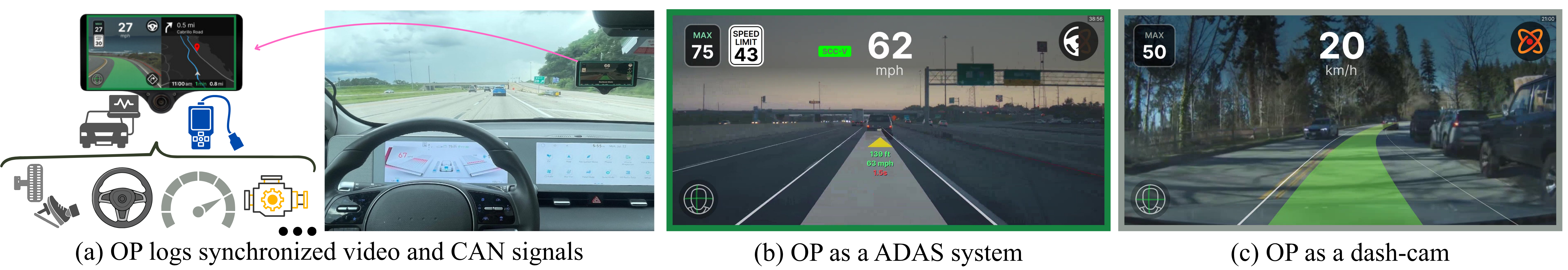}
    \caption{Data-collection configurations in ADAS-TO. The openpilot system either controls the vehicle or records synchronized video and signals while the vehicle's OEM ADAS is active.}
    \label{fig:collection_setup}
\end{figure}

We reconstruct automation engagement from drive-level logs and identify engaged-to-disengaged transitions. These transitions define the dataset events, referred to as takeovers below; the recorded boundary need not coincide with the first driver input \citep{10285570}.
Temporal validity constraints suppress artifacts: minimum engaged and disengaged dwell times around each transition, bridging of sub-second engagement gaps, and debouncing of repeated transitions.

\begin{figure}[pos=htbp]
    \centering
    \includegraphics[width=0.55\linewidth]{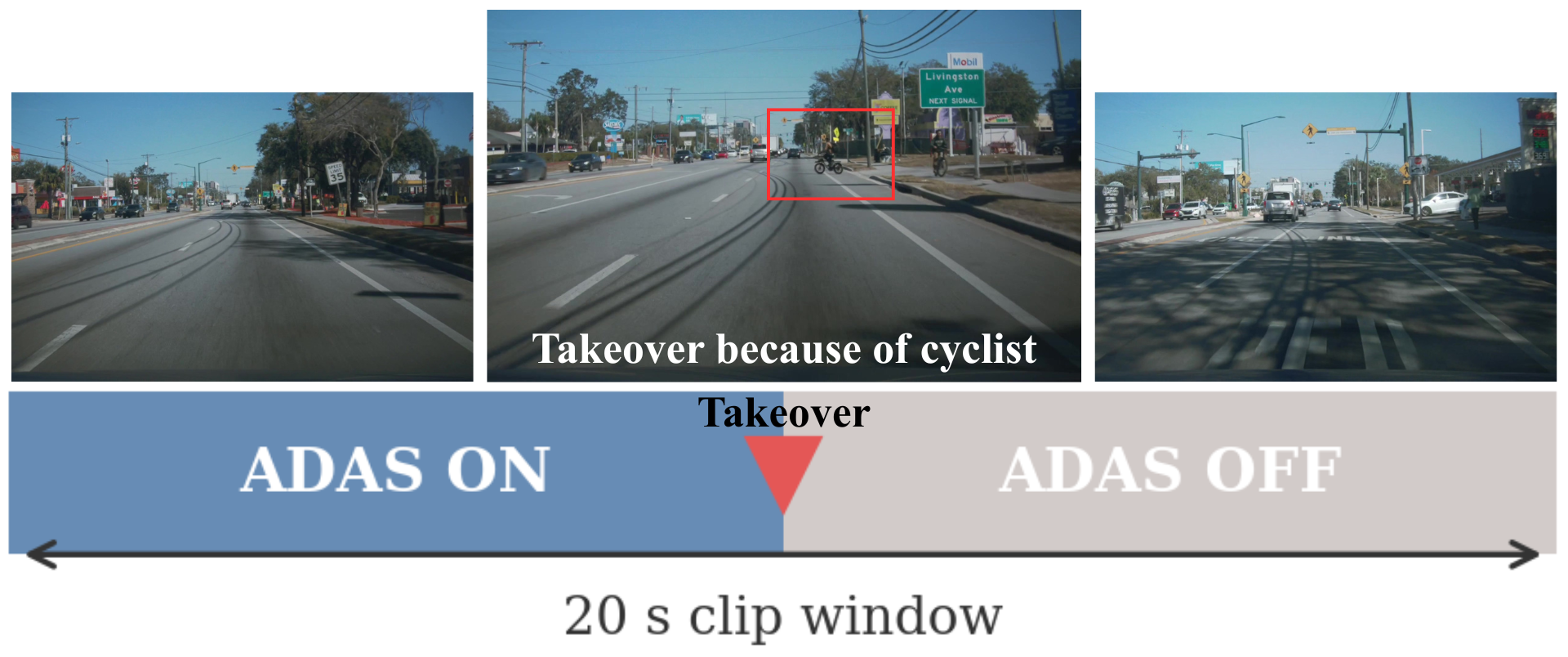}
    \caption{ADAS-TO clip structure. Each 20-s clip spans 10~s before and 10~s after logged disengagement, which occurs 10~s into the clip ($t=0$ in the analysis). Driver inputs can begin before this edge, and later automation-state changes remain in the record.}
    \label{fig:ADAS_TO_structure}
\end{figure}

Each takeover yields a 20~s clip centered on the transition, covering the preceding approach and subsequent vehicle and driver response.
Clips are self-contained multimodal records (front video, metadata, synchronized vehicle, automation, perception, radar, and inertial signals; Table~\ref{tab:clip_files}).

\subsection{Dataset Overview}
\label{sec:dataset_overview}

\begin{figure}[pos=!htbp]
    \centering
    \includegraphics[width=\linewidth]{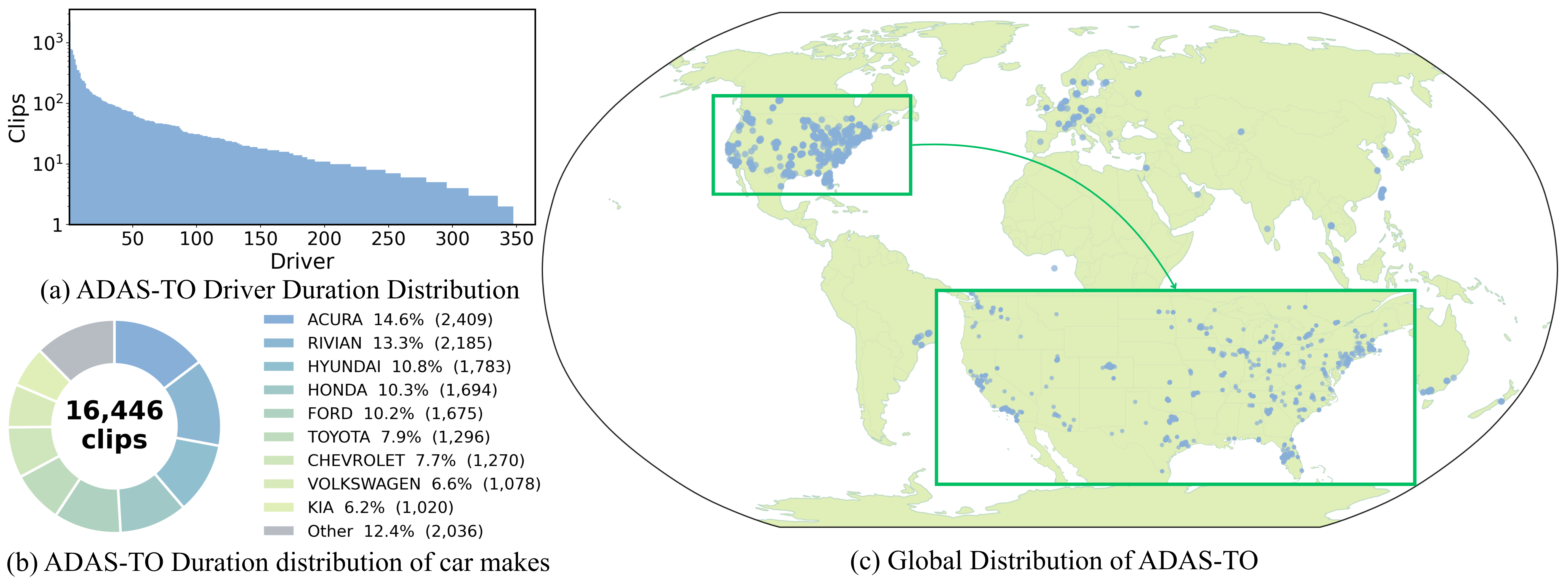}
    \caption{ADAS-TO coverage by driver, vehicle make, and location.}
    \label{fig:dataset_distribution}
\end{figure}

ADAS-TO contains 16,446 takeover clips from 364 different drivers across 2,585 different drives.
It covers 99 vehicle models from 23 makes, after combining model years, generations, and powertrain variants within the same model family.
The grouping is listed in Appendix~\ref{app:vehicle_models}.
Each clip contains 20~s of video at 20~fps and synchronized signals recorded at up to 100~Hz.
Both openpilot-controlled and factory-ADAS driving are represented, with factory systems accounting for about two-thirds of the clips.
Most recordings are from North America, with additional contributions from other countries (Fig.~\ref{fig:dataset_distribution}).
All model evaluation separates drivers between training and testing.

\begin{table}[pos=htbp]
\centering
\caption{Data components released for each takeover clip (15 files: one video, one metadata file, and 13 CSV files).}
\label{tab:clip_files}
\small
\setlength{\tabcolsep}{4pt}
\renewcommand{\arraystretch}{1.15}
\begin{tabularx}{\textwidth}{
>{\raggedright\arraybackslash}p{0.20\textwidth}
>{\raggedright\arraybackslash}p{0.27\textwidth}
>{\raggedright\arraybackslash}X}
\toprule
Component & Files & Main content and analytical use \\
\midrule
Front-view video &
\texttt{takeover.mp4} &
20~s front-view clip centered on the takeover instant; supports visual scene understanding and semantic hazard annotation. \\
Metadata &
\texttt{meta.json} &
Vehicle, driver, drive, and clip identifiers, source log/video type, sampling rate, takeover timestamp, clip start time, and source segment information. \\
Ego vehicle and driver input signals &
\texttt{carState.csv}; \texttt{accelerometer.csv} &
Speed, acceleration, steering angle/torque, gas and brake inputs, cruise state, and IMU acceleration; supports longitudinal and lateral response analysis. \\
Inertial, odometry, and calibration signals &
\texttt{VehicleIMU.csv}; \texttt{Gyroscope.csv}; \texttt{CameraOdometry.csv}; \texttt{LiveCalibration.csv}; \texttt{LiveParameters.csv} &
Body-frame IMU and yaw rate, angular rates, visual odometry, camera-to-vehicle calibration, and online vehicle-parameter estimates; supports post-takeover motion analysis. \\
Automation control and planner signals &
\texttt{carControl.csv}; \texttt{carOutput.csv}; \texttt{controlsState.csv}; \texttt{longitudinalPlan.csv} &
System activation, lateral/longitudinal control states, commanded acceleration, torque, speed, curvature, target acceleration, planned profiles, and alert states; supports analysis of automation behavior before intervention. \\
Perception and surrounding-traffic signals &
\texttt{drivingModelData.csv}; \texttt{radarState.csv} &
Lane-line probabilities and offsets, desired curvature, lead-vehicle status, relative distance, relative velocity, and relative acceleration; supports lane-keeping, car-following, TTC, and THW analysis. \\
\bottomrule
\end{tabularx}
\end{table}

\begin{figure}[pos=htbp]
    \centering
    \includegraphics[width=\linewidth]{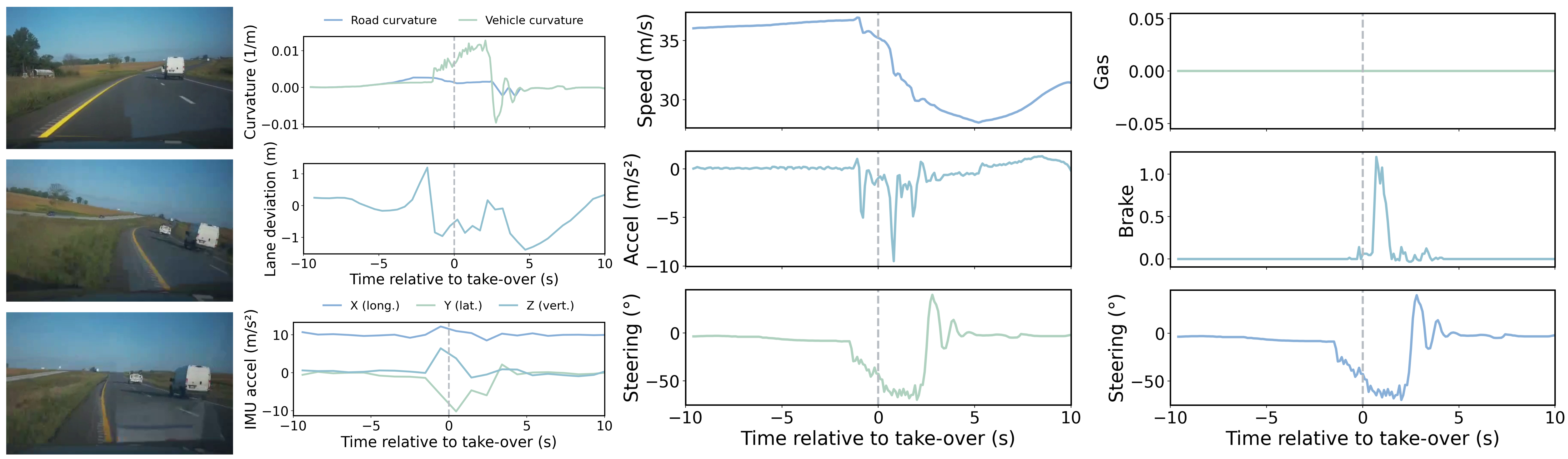}
    \caption{Multi-modal data example in ADAS-TO, with aligned video and CAN signals.}
    \label{fig:dataset_example}
\end{figure}

The synchronized video and signals allow vehicle motion and driver inputs to be examined in their road context (Fig.~\ref{fig:dataset_example}).
Longitudinal measures include speed, acceleration, headway, and TTC; lateral measures include steering, lane confidence, and lane-relative position.
Signal availability and meaning depend on the vehicle and recording configuration.
Lead-based measures use the recorded lead representation; internal OEM target-selection states are unavailable.

\subsection{Manually Labeled Takeover Events}
\label{sec:dataset_filtering}
\label{sec:maneuver_filtered_adas}

We manually identified 6,206 takeover events during continued lane keeping or car following, referred to as the Maneuver-Filtered ADAS subset in the analyses.
Signal-based rules first flagged turns, lane changes, and traffic-control stops.
Reviewers then examined the full video and synchronized signals to distinguish these maneuvers from interventions during otherwise continuous driving.
Events with unclear road context or a separate maneuver were excluded from the analysis.
The review criteria and scenario breakdown are provided in Appendix~\ref{app:maneuver_filtered_subset}.

The selected events cover both urban and higher-speed driving.
Low-speed following was retained when the driver remained behind a lead vehicle without approaching a traffic-control stop.
This selection allows us to examine braking and steering during takeover without mixing in the actions required for a turn, lane change, or intersection stop.
The descriptive analysis follows these events before and after disengagement.

\subsection{Car-Following Samples for Modeling}
\label{sec:methods:ccdataset}

For the response model, we use takeovers in which the ego vehicle is closing on a tracked lead.
The reference is calibrated using 822 cases that continue following the same lead for at least five seconds after takeover.
To study disengagement, we compare closing episodes that contain a takeover with episodes in which ADAS remains active.
Comparison episodes are sampled from the same drives, with similar starting speeds and no nearby disengagement.
The primary model uses episodes with no driver input before closing begins.
Section~\ref{sec:methods:hazard} defines the model and follow-up period; the detailed sampling rules and counts are in Appendix~\ref{app:robustness}.

\section{Driver Intervention and Longitudinal Conflict Dynamics}
\label{sec:stats}
\label{takeover_analysis}

We analyze the 6,206 manually filtered takeover events to describe when drivers intervene and how vehicle motion changes afterward.
The analysis follows driver inputs, TTC, braking, and conflict resolution within the available follow-up.
We also review the road scenes in safety-critical cases to understand the situations in which these interventions occur.

\subsection{Event Alignment and Measures}
\label{sec:methods:events_framework}
\label{sec:methods:measures_short}

We set $t=0$ when ADAS switches from engaged to disengaged, at the center of the clip.
The principal descriptive window is $[-5,+5]$~s.
For events with a brake or steering action, $T_{\mathrm{input}}$ denotes the onset of the first such action, and intervention lead time is
\begin{equation}
L_{\mathrm{act}}=T_{\mathrm{dis}}-T_{\mathrm{input}}.
\label{eq:lact}
\end{equation}
Positive values indicate action before disengagement; zero and negative values indicate action at or after it.
Brake and accelerator activity are obtained from pedal signals.
Steering is active when a driver press is detected together with an absolute wheel rate above $2^\circ$/s.
Accelerator input is retained as a response channel but excluded from $T_{\mathrm{input}}$, because its override and disengagement behavior differs across platforms.

Longitudinal measures include spacing $s$, closing speed $c=v_E-v_L$, and $\mathrm{TTC}=s/c$ when $c>0$.
We represent looming using the kinematic measure $\Lambda=\max(c,0)/s$, which equals inverse TTC during closing and approaches zero as closing ceases.
A larger looming value means that the gap is closing more quickly relative to the distance available.
Pre/post comparisons use extrema over equal-duration windows, $[-5,0)$ and $[0,+5]$~s.
Paired analyses additionally require valid measurements in both windows.
Appendix~\ref{app:measures} gives the signal-validity rules and remaining measure definitions.

\subsection{Driver-Action Timing and Control Channels}
\label{sec:timing_modality}

\begin{figure}[pos=htbp]
\centering
\includegraphics[width=0.6\linewidth]{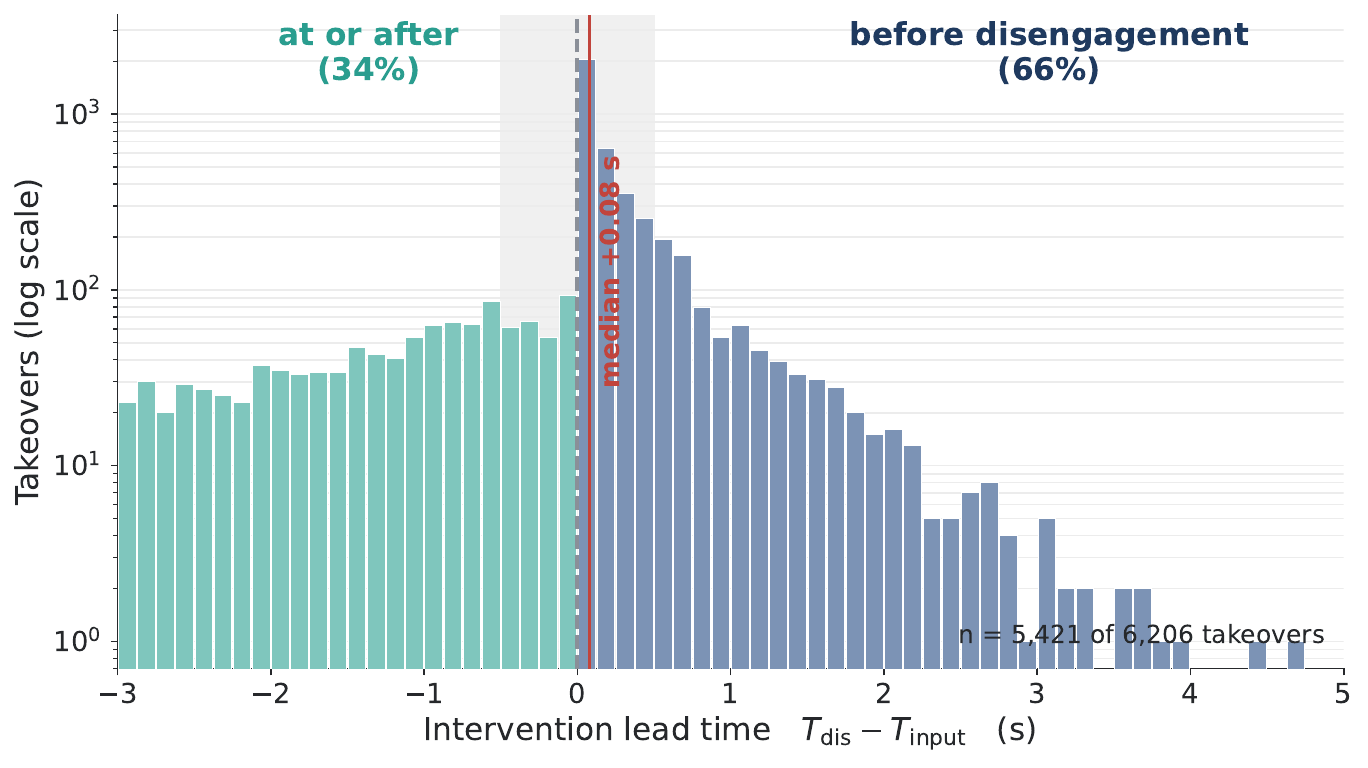}
\caption{Timing of the first brake or steering action relative to ADAS disengagement. Positive values mean that the driver started intervening while ADAS was still active; zero or negative values mean that the action began at or after disengagement. The histogram includes takeovers with a recorded brake or steering action. Counts use a logarithmic scale.}
\label{fig:takeover_timeliness}
\end{figure}

Braking or steering was recorded in most takeover events.
In 65.6\% of these cases, the driver began pressing the brake or turning the steering wheel while ADAS was still active, and the system disengaged afterward.
In the remaining cases, these actions began at or after disengagement.
The typical separation is short---the median lead is 0.08~s---but the sequence matters: the driver's intervention can already be underway before the recorded transition to manual driving (Fig.~\ref{fig:takeover_timeliness}).
The recorded ordering reflects both driver actions and the system's input and disengagement logic.

\begin{figure}[pos=htbp]
\centering
\includegraphics[width=\linewidth]{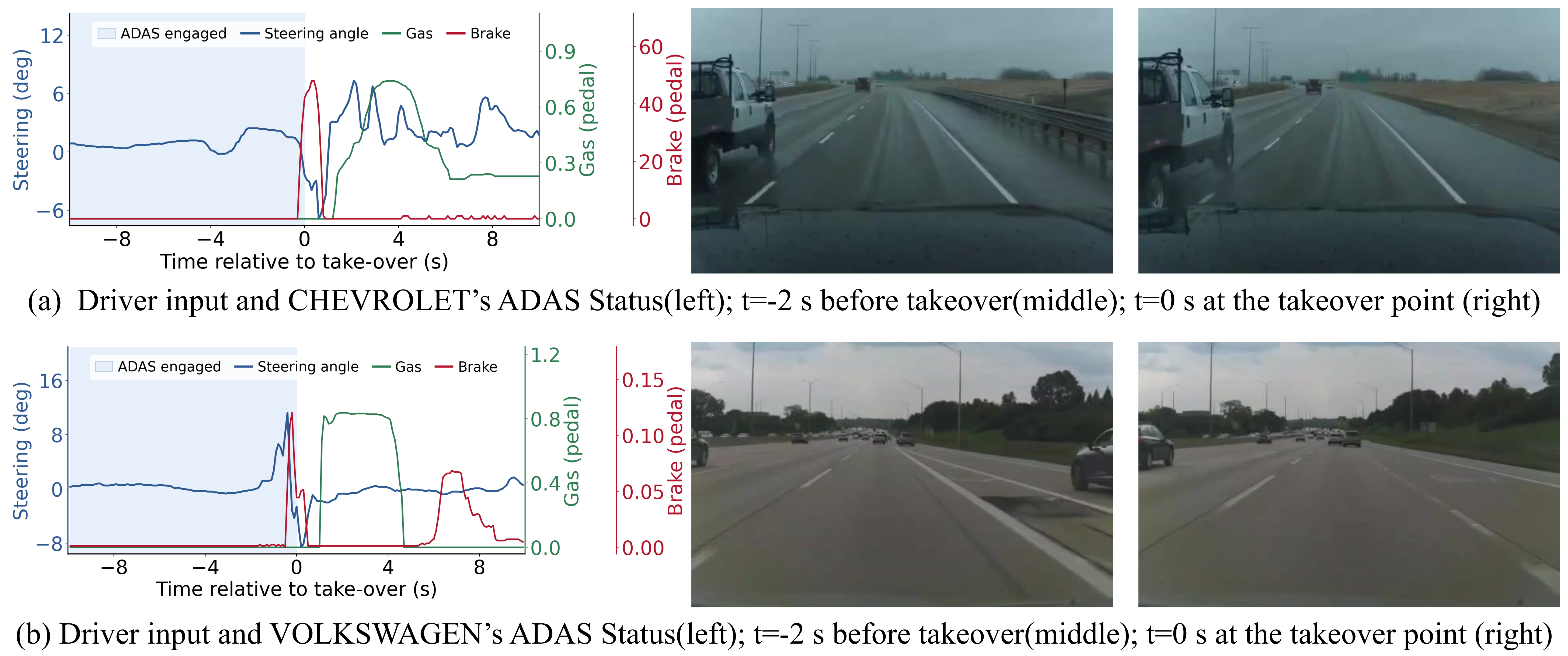}
\caption{
Examples of driver input and ADAS engagement state around takeover.
The left panels show steering, gas, brake, and engagement status.
The middle and right panels show the road scene before and at the takeover point.
}
\label{fig:system_status}
\end{figure}

The examples in Fig.~\ref{fig:system_status} illustrate why the two timestamps can differ.
Braking commonly disengages longitudinal assistance, accelerator input may override or disengage it, and a button or stalk cancellation can end engagement without a pedal or steering signature.
The subsequent input profiles also differ by channel.
Braking tends to peak near the transfer and then decline, whereas steering remains active for longer and accelerator input becomes more common later in the observation window (Appendix~\ref{app:patterns}).

\subsection{Evolution of Longitudinal Safety Margins}
\label{sec:longitudinal_takeover_dynamics}

\begin{figure*}[pos=htbp]
\centering
\includegraphics[width=0.75\textwidth]{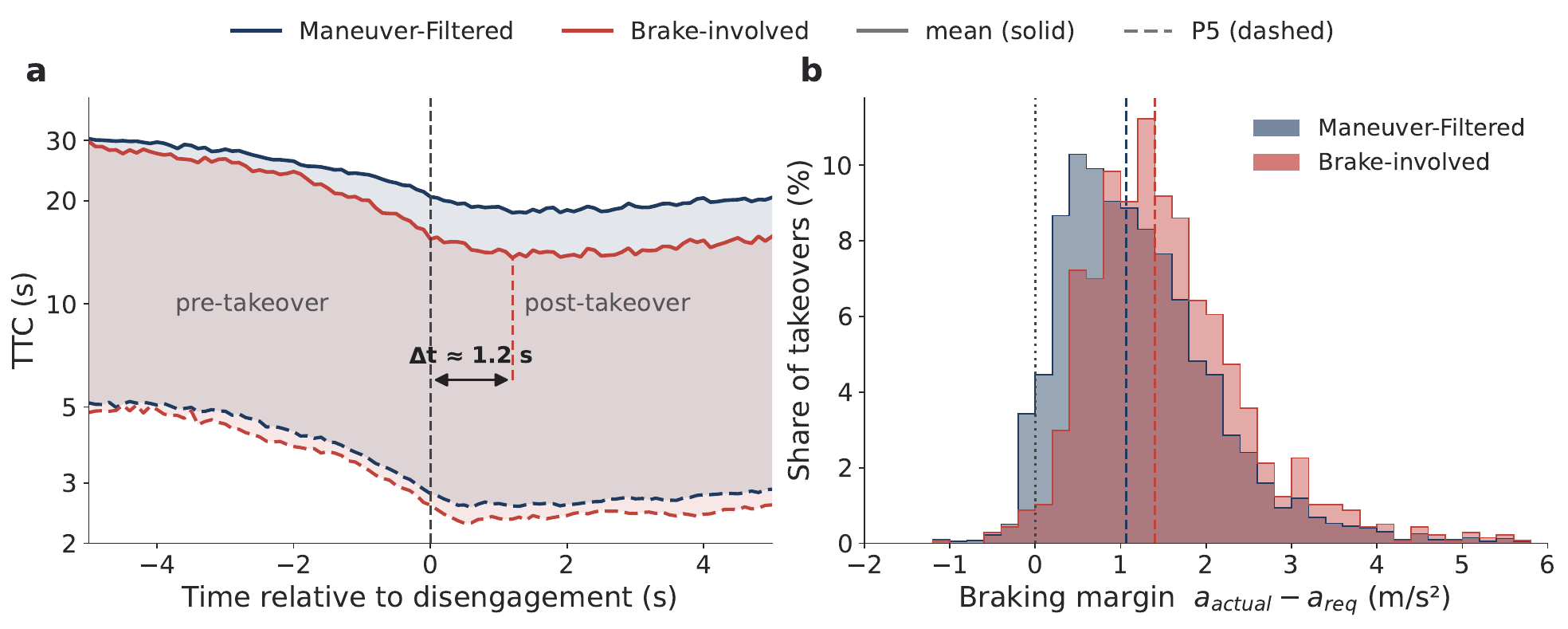}
\caption{
Longitudinal TTC and braking margin around ADAS disengagement.
Panel~(a) shows the mean (solid) and fifth-percentile (dashed) TTC trajectories for the Maneuver-Filtered ADAS subset and the brake-involved subset.
The vertical dashed line marks disengagement, and the red marker indicates the approximate time of the minimum mean TTC in brake-involved events.
Panel~(b) shows the distribution of realized braking margin, $a_{\mathrm{actual}}-a_{\mathrm{req}}$, in valid closing-lead cases.
The dotted line denotes zero braking margin, and colored dashed lines denote group medians.
}
\label{fig:longitudinal_transit}
\end{figure*}

Figure~\ref{fig:longitudinal_transit}a shows the mean and fifth-percentile TTC trajectories.
The mean curves decline before disengagement and continue to fall afterward.
For brake-involved events, mean TTC reaches its minimum about 1.2~s after disengagement; the lower-tail trajectory has a similar shape.
This timing describes the minimum of the event-aligned mean curve, rather than the average time of individual TTC minima.
It shows that the aggregate TTC trajectory continues to deteriorate after the change in control source.

\begin{figure*}[pos=htbp]
\centering
\includegraphics[width=0.8\textwidth]{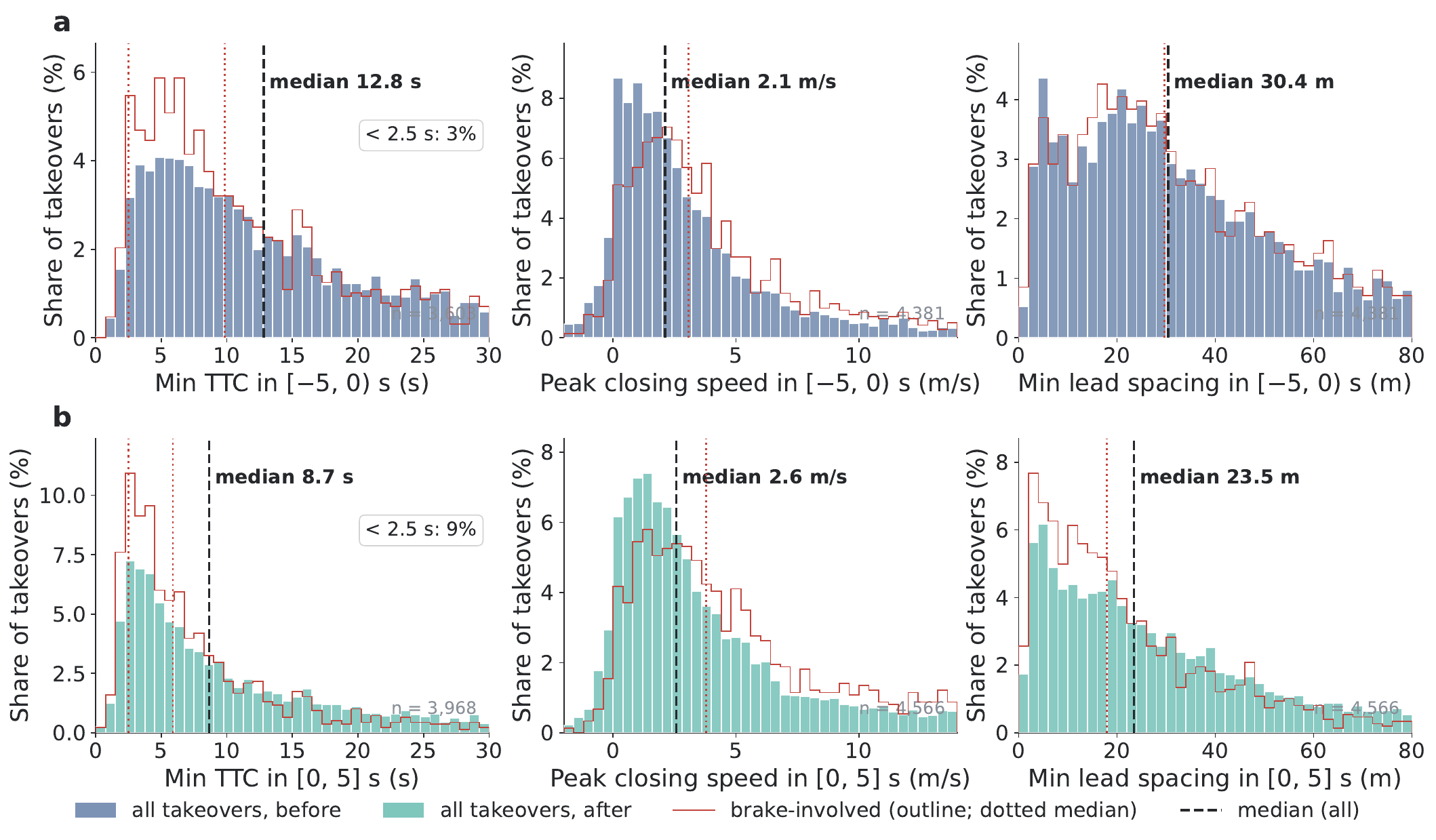}
\caption{
Longitudinal extrema in symmetric windows before and after the disengagement edge, for all aligned Maneuver-Filtered takeovers with a valid closing lead.
The upper row shows the minimum TTC, peak closing speed, and minimum lead spacing over $[-5,0)$~s; the lower row shows the same quantities over $[0,+5]$~s.
Filled bars are all takeovers; red outlines are brake-involved takeovers.
Dashed lines denote medians (dotted red for brake-involved), and the red dotted line in the TTC panels denotes the 2.5-s low-margin threshold.
Each panel reports the number of takeovers with a valid measurement in that window.
}
\label{fig:longitudinal_distribution}
\end{figure*}

The distributions before and after takeover show the same tendency (Fig.~\ref{fig:longitudinal_distribution}).
The median minimum TTC decreases from 12.8 to 8.7~s, with a larger share of low-TTC events after takeover.
The change is more pronounced in cases involving braking.
Closing speed also tends to increase and spacing to decrease during this period.
The figure reports the available sample for each measure.

Among 2,031 car-following cases in which the ego vehicle was still approaching the lead at takeover, 78.5\% reached a TTC within the next five seconds that was below their minimum in the preceding five seconds.
We observe a similar proportion when the same lead remains tracked throughout the following five seconds (Table~\ref{tab:post_summary}).
These paired comparisons show that a lower TTC often occurs after disengagement, even though the driver has begun responding.

\authorcheck{D2}{The figure and paired table use different validity filters. The current prose distinguishes their populations. Reconcile both extraction paths before reporting a single pooled change or denominator.}

\begin{table}[pos=htbp]
\centering
\caption{Minimum TTC before and after takeover and estimated time to conflict resolution. A lower post-takeover minimum means that minimum TTC in $[0,5]$~s is below its minimum in $[-5,0)$~s. Recovery columns are Kaplan--Meier estimates under non-informative censoring; brackets give driver-clustered uncertainty ranges. Follow-up is described in Appendix~\ref{app:recovery_followup}.}
\label{tab:post_summary}
\small
\setlength{\tabcolsep}{4pt}
\renewcommand{\arraystretch}{1.15}
\begin{tabularx}{\linewidth}{@{}Xrccc@{}}
\toprule
Sample & $N$ & \shortstack{Lower post-takeover\\minimum TTC} & \shortstack{Median resolution\\time (s)} & \shortstack{Resolved\\within 5 s} \\
\midrule
All paired events & 3,827 & \shortstack{62.5\%\\{[59.1, 65.5]}} & \shortstack{0.92\\{[0.61, 1.17]}} & \shortstack{81.9\%\\{[79.6, 84.4]}} \\
Closing at disengagement & 2,031 & \shortstack{78.5\%\\{[75.6, 80.9]}} & \shortstack{2.11\\{[1.86, 2.38]}} & \shortstack{75.5\%\\{[71.8, 79.2]}} \\
Closing, same lead for 5 s & 994 & \shortstack{78.4\%\\{[74.0, 81.8]}} & \shortstack{1.92\\{[1.61, 2.19]}} & \shortstack{78.0\%\\{[74.2, 81.5]}} \\
\bottomrule
\end{tabularx}
\end{table}

\subsection{Braking Response and Conflict Resolution}
\label{sec:braking_recovery}

Observed peak deceleration generally exceeds the instantaneous kinematic braking demand, particularly in brake-involved events (Fig.~\ref{fig:longitudinal_transit}b).
The demand reference assumes unchanged lead speed and no response delay, so this margin describes the relation between the recorded response and the closing geometry.

\begin{figure*}[pos=htbp]
\centering
\includegraphics[width=0.75\textwidth]{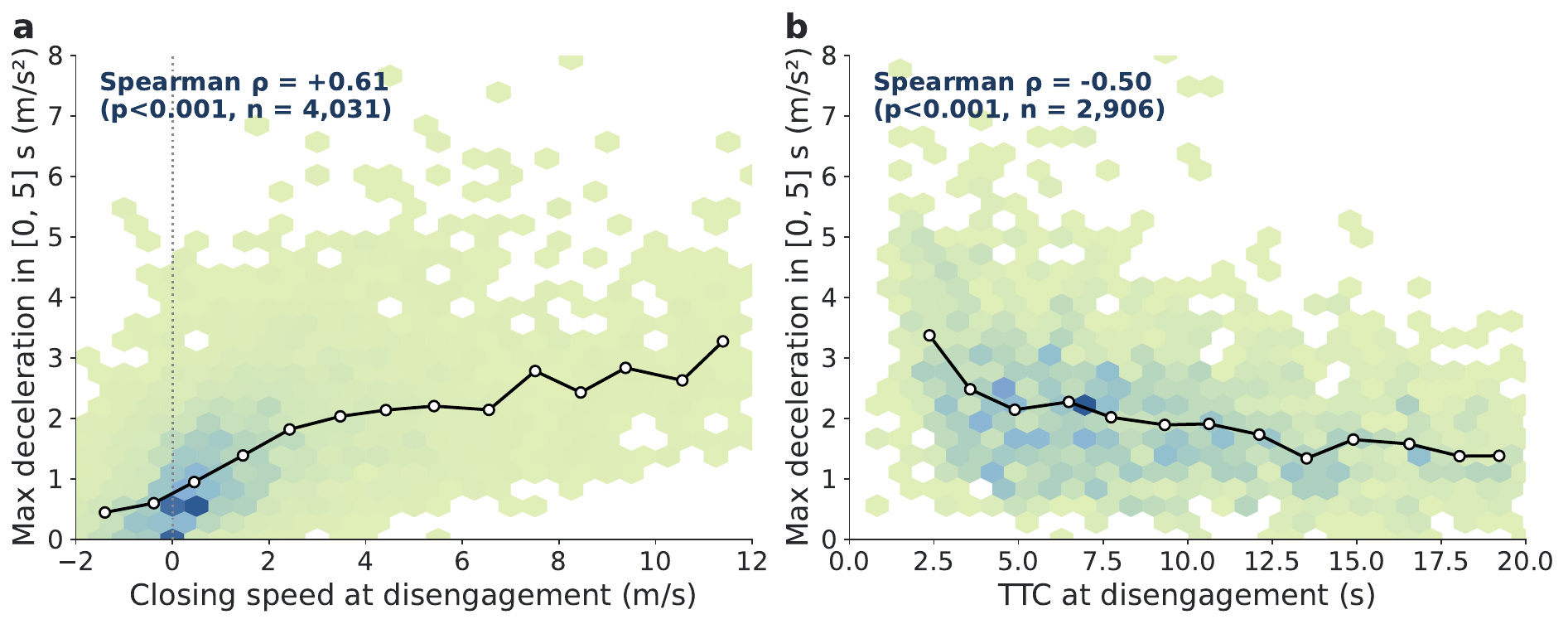}
\caption{
Associations between maximum post-takeover deceleration and longitudinal traffic conditions in valid closing-lead cases.
Panel~(a) relates maximum deceleration to relative speed at disengagement.
Panel~(b) relates maximum deceleration to TTC at disengagement.
Black lines show binned median deceleration, and the reported Spearman correlations summarize rank associations over all valid cases; axes are truncated for display, and peaks above 9.8~m/s$^2$ are excluded by the signal-quality screen.
}
\label{fig:brake_dynamics}
\end{figure*}

Drivers brake more strongly when they are approaching the lead faster or have less time before a potential collision.
Peak deceleration is positively associated with closing speed ($\rho=0.61$) and negatively associated with TTC at takeover ($\rho=-0.50$; Fig.~\ref{fig:brake_dynamics}).
The comparisons use the valid observations for each variable, so their correlation magnitudes should not be read as a direct test of which cue drivers rely on more.

We also examine how long it takes for the closing conflict to resolve.
Resolution requires TTC to return to its pre-takeover reference, or the ego vehicle to stop approaching the lead, for at least one second.
Kaplan--Meier estimates give a median resolution time of about 2.1~s and a five-second resolution probability of roughly three-quarters for the closing cases (Table~\ref{tab:post_summary}).
The same-lead analysis gives a similar result.
These estimates assume that events ending observation early have comparable remaining resolution times to those still observed.
Lead loss may depend on the traffic situation, so the estimates remain conditional on this assumption; the follow-up counts and interpretation are given in Appendix~\ref{app:recovery_followup}.
This recovery measure describes the return to the earlier longitudinal condition or the end of closing, and differs from the time of minimum TTC.

\begin{table}[pos=htbp]
\centering
\caption{Longitudinal contexts in the manually reviewed safety-critical set ($n=749$). Labels can overlap. Lead-representation discrepancy is a subtype of rapid closing.}
\label{tab:discussion_context_distribution}
\small
\begin{tabularx}{\linewidth}{@{}Xrr@{}}
\toprule
Context & Events & Share of review set \\
\midrule
Rapid closing toward an in-path lead & 323 & 43.1\% \\
\quad Apparent lead-representation discrepancy & 97 & 13.0\% \\
Sudden cut-in & 82 & 10.9\% \\
Danger sign or warning context & 16 & 2.1\% \\
\bottomrule
\end{tabularx}
\end{table}

\subsection{Longitudinal Contexts in Safety-Critical Events}
\label{sec:critical}
\label{sec:discussion_rapid_closing}

We reviewed 749 safety-critical cases to identify the traffic situations surrounding takeover.
The screening uses indicators of low safety margins, unusual vehicle motion, or strong driver response; details are provided in Appendix~\ref{app:safety_relevant_cases}.
This separate review set helps interpret the patterns in the descriptive analysis.

Rapid closing is the most common longitudinal context, accounting for about 43\% of reviewed cases.
Cut-ins are another recurring situation.
Some rapid-closing cases also show a discrepancy between the lead vehicle visible in the video and the lead represented in the recorded signals.
Table~\ref{tab:discussion_context_distribution} summarizes these labels, and Table~\ref{tab:all_review_contexts} gives the full inventory.
Here rapid closing is a scene label assigned during manual review; the larger closing sample used above includes all eligible cases in which the ego vehicle is approaching a lead.

\begin{figure*}[pos=htbp]
\centering
\includegraphics[width=0.98\textwidth]{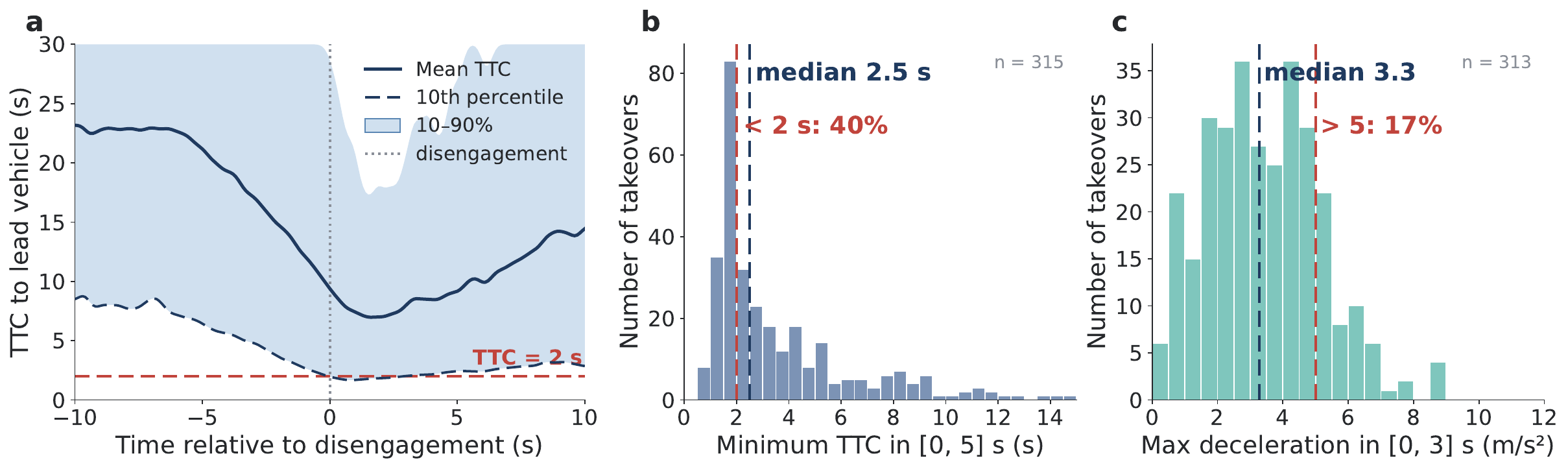}
\caption{Longitudinal conditions in rapid-closing events.
Mean TTC declines toward the takeover and remains low shortly afterward.
The post-takeover minimum TTC has a median of 2.5~s, with 40\% of events below 2~s, while maximum deceleration over $[0,+3]$~s has a median of 3.3~m/s$^2$ (peaks above 9.8~m/s$^2$ excluded by the signal-quality screen).}
\label{fig:overall_rapid_closing}
\end{figure*}

In these rapid-closing cases, TTC decreases before takeover and remains low shortly afterward (Fig.~\ref{fig:overall_rapid_closing}).
Forty percent have a post-takeover TTC below 2~s.
The subsequent braking is often strong, with 17\% exceeding 5~m/s$^2$ after the signal-quality screen.
The examples therefore show why the period after takeover matters: even an active braking response can occur while the vehicles are still getting closer.

\begin{figure*}[pos=htbp]
\centering
\includegraphics[width=0.9\textwidth]{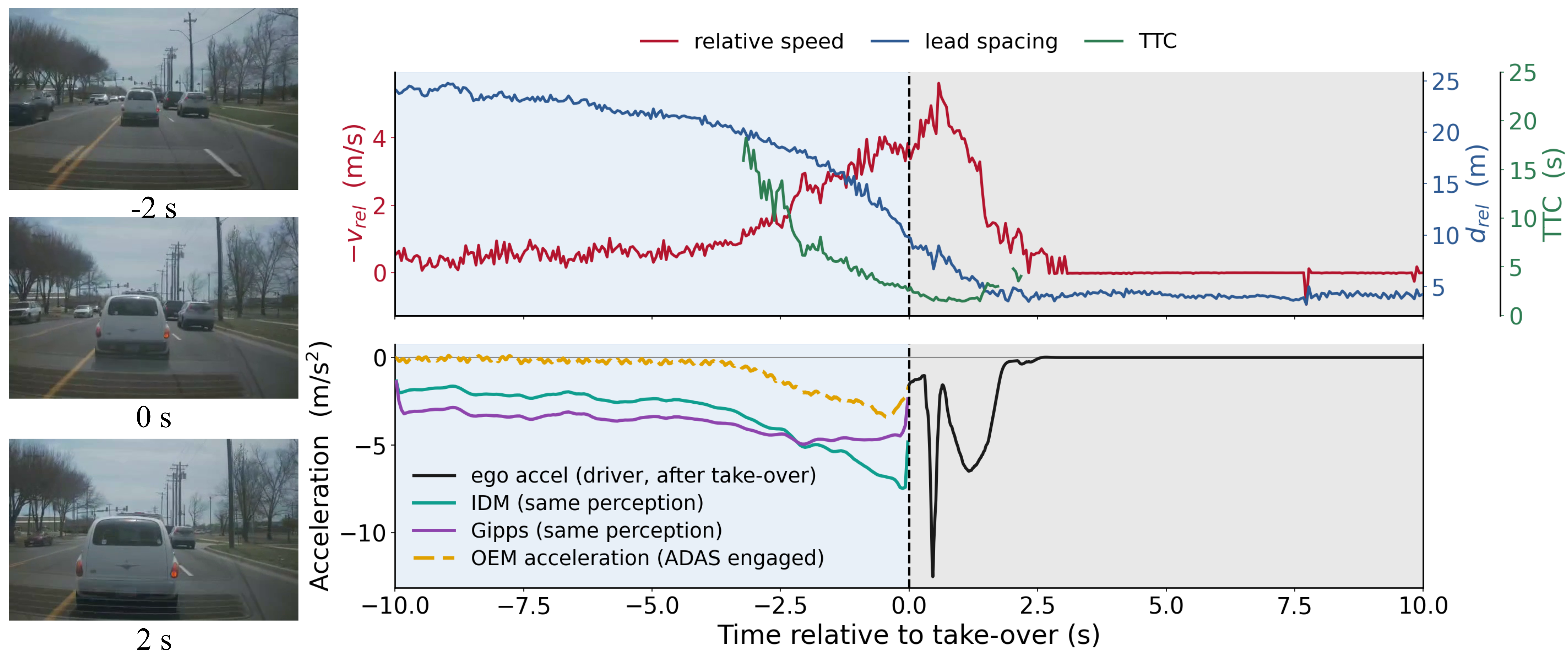}
\caption{Representative rapid-closing takeover.
The upper panel shows increasing relative speed, decreasing lead spacing, and declining TTC before takeover.
The lower panel compares logged vehicle acceleration during OEM ADAS engagement with IDM and Gipps reference-model accelerations computed from the same smoothed radar-derived lead state.
After takeover, the black curve shows the observed human-driven vehicle acceleration.}
\label{fig:high_relative_speed}
\end{figure*}

The event in Fig.~\ref{fig:high_relative_speed} illustrates the sequence.
Vehicle acceleration during OEM assistance remains near zero through much of the approach and becomes negative late; the driver then brakes sharply.
IDM and Gipps references supplied with the same logged lead state begin decelerating earlier.
This comparison describes the timing of the recorded response relative to two model references under the available lead representation.

\begin{figure*}[pos=htbp]
\centering
\includegraphics[width=0.8\textwidth]{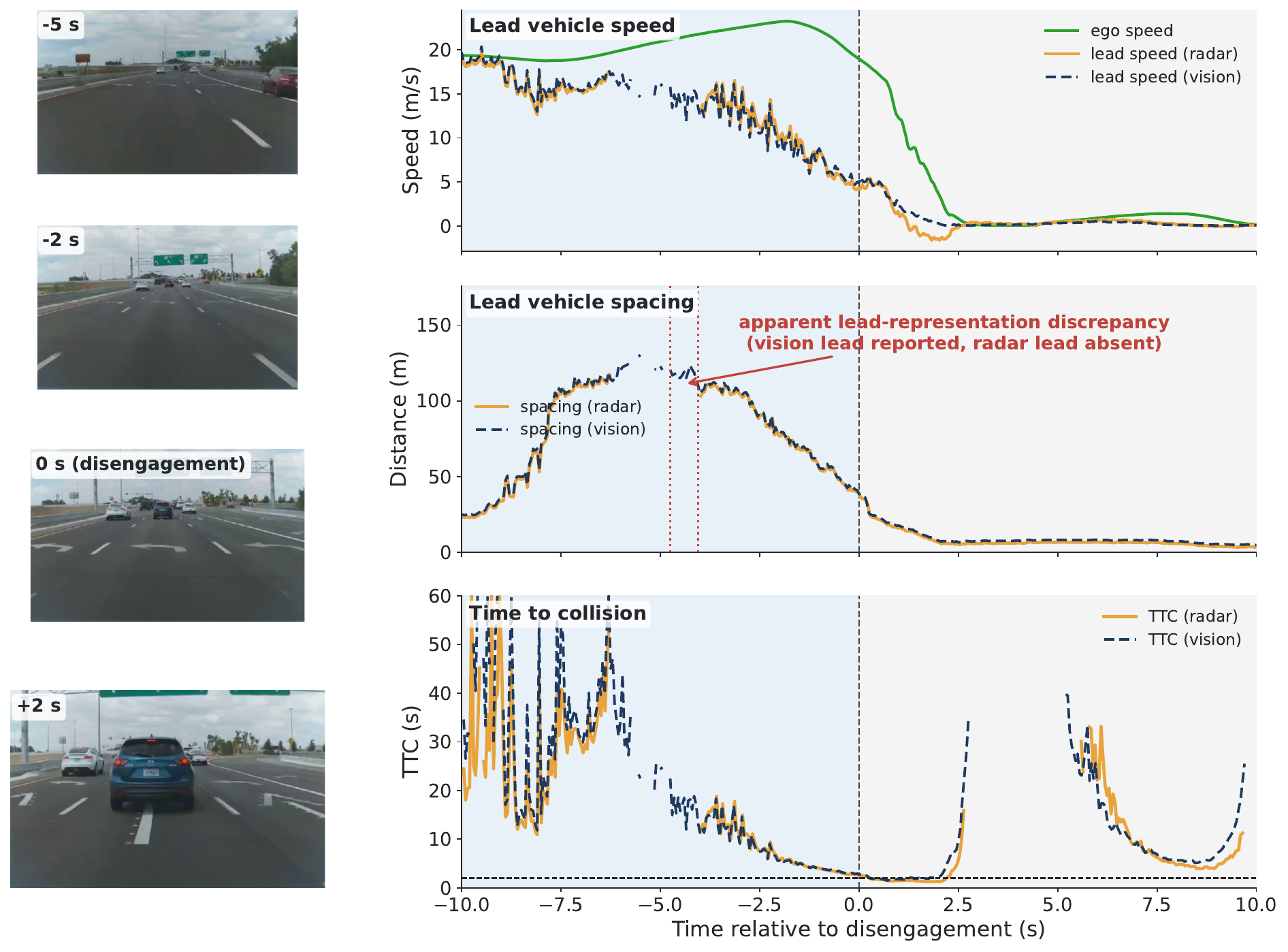}
\caption{Example of a rapid-closing takeover with an apparent lead-representation discrepancy.}
\label{fig:miss_radar}
\end{figure*}

About 30\% of the rapid-closing cases show an apparent lead-representation discrepancy.
In these records, a visually relevant in-path lead is absent from, delayed in, or inconsistent with the logged lead state (Fig.~\ref{fig:miss_radar}).
The recordings establish a discrepancy between the available video and signal evidence; they do not identify a particular sensor, tracker, or OEM controller as its cause.
\authorcheck{D3}{Reconcile the 749-case review set and its screening denominators with the revised scenario labels, including the 14 road-disturbance labels. The current version treats it as a separate review set and preserves the supplied context counts.}

\section{Looming-Based Modeling of Braking and ADAS Disengagement}
\label{sec:model}

The preceding results show that longitudinal conflicts can persist through the control transfer and that drivers adjust their braking to the approach.
We now use the observed manual response to define a reference for the behavior before takeover.
GHR is used for its compact relation between looming and braking, with speed and spacing modifying the response.
The analysis has two parts: a GHR law is fitted to post-takeover following trajectories, and departures from this fitted response are related to the occurrence of disengagement.

\subsection{Post-Takeover Response Model}
\label{sec:methods:benchmark}

The calibration sample contains 822 cases with at least 5~s of following the same lead after disengagement, denoted by $t_D$.
We represent the longitudinal response with the GHR law \citep{gazis1961follow},
\begin{equation}
a_H^{\mathrm{cmd}}(t)=\gamma v(t)^m
\frac{v_L(t-\tau)-v(t-\tau)}{s(t-\tau)^l},
\label{eq:ghr}
\end{equation}
where $v$ and $v_L$ are ego and lead speed, $s$ is spacing, $\gamma$ scales the response, $m$ and $l$ govern its speed and spacing dependence, and $\tau$ is an effective delay.
A shared vehicle-response layer applies a first-order lag of 0.3~s and a jerk limit before acceleration is integrated to obtain speed and spacing.
The fitted law describes the average response represented by these post-takeover trajectories.

Calibration and evaluation use five folds with different drivers in the training and test sets.
Each test driver's trajectories are simulated using parameters fitted on the other drivers.
IDM \citep{treiber2000idm} and Gipps \citep{gipps1981carfollowing} are calibrated under the same protocol.
The evaluation distinguishes two lead inputs.
The conditional evaluation supplies the lead motion observed during the post-takeover interval and assesses the follower response given that motion.
The ex-ante evaluation extrapolates the lead's constant-acceleration trend from information available at $t_D$, so its error also includes uncertainty about the subsequent lead trajectory.
In both cases, ego speed and spacing are rolled forward rather than replaced by their measured values at each step.
\authorcheck{M1}{Verify the 822-event calibration cohort against the final labels, and provide the fitted objective, parameter bounds, integration step, jerk limit, and preprocessing settings. These implementation details were not included in the supplied project.}

\subsection{Trajectory Fit and Physical Interpretation}
\label{sec:model:post}
\label{sec:model:looming}

The GHR model reproduces post-takeover following trajectories with errors close to IDM and smaller than Gipps (Table~\ref{tab:ghr_holdout}).
Given the measured lead trajectory, its median event-level five-second spacing RMSE is 1.34~m.
When the lead's motion must also be predicted, the error increases to 1.84~m.
The corresponding speed errors are about 0.55 and 0.90~m/s.
The difference between the two evaluations shows how uncertainty about the lead vehicle affects the prediction, even when the model of the following vehicle is unchanged.

\begin{table}[pos=htbp]
\centering
\caption{Post-takeover trajectory errors evaluated on held-out drivers (822 events). Entries are medians of event-level RMSE over the indicated rollout horizon. Conditional evaluation uses observed lead motion; ex-ante evaluation extrapolates it at disengagement.}
\label{tab:ghr_holdout}
\small
\setlength{\tabcolsep}{5pt}
\begin{tabular}{llcccc}
\toprule
Lead input & Model & \multicolumn{3}{c}{Spacing RMSE (m)} & Speed RMSE (m/s) \\
\cmidrule(lr){3-5}
 & & 1~s & 3~s & 5~s & 5~s \\
\midrule
Conditional & GHR & 0.353 & 0.831 & 1.342 & 0.547 \\
 & IDM & 0.354 & 0.873 & 1.348 & 0.579 \\
 & Gipps & 0.360 & 0.984 & 1.494 & 0.647 \\
\midrule
Ex-ante & GHR & 0.386 & 0.993 & 1.837 & 0.901 \\
 & IDM & 0.392 & 1.056 & 1.837 & 0.981 \\
 & Gipps & 0.394 & 1.112 & 2.000 & 0.980 \\
\bottomrule
\end{tabular}
\end{table}

The full-sample parameter estimates are $\gamma=3.21$, $m=0.74$, $l=1.22$, and $\tau=0.34$~s.
During closing, $c=s\Lambda$, so Eq.~\eqref{eq:ghr} can be written as
\begin{equation}
\begin{aligned}
a_H^{\mathrm{cmd}}(t)
&=-\gamma v(t)^m s(t-\tau)^{1-l}\Lambda(t-\tau)\\
&=-3.21\,v(t)^{0.74}s(t-\tau)^{-0.22}\Lambda(t-\tau).
\end{aligned}
\label{eq:ghr_loom}
\end{equation}
Equation~\eqref{eq:ghr_loom} gives the fitted response a practical interpretation in terms of looming.
Looming increases when the ego vehicle approaches the lead faster or when the remaining gap becomes shorter.
Within the fitted law, doubling looming at a given speed and spacing doubles the braking command.
The positive speed exponent gives a stronger response to the same looming at higher driving speeds.
The small negative spacing exponent adds a further increase when the gap is shorter, even at the same looming.
In this way, the model describes a response to how quickly the available space is being used up \citep{lee1976theory,markkula2016farewell}.

The linear dependence on looming at a given speed and spacing is part of the GHR form; calibration estimates the response gain and its speed, spacing, and delay parameters.
The trajectory results support this relationship as a compact description of post-takeover braking across drivers.
They do not identify looming as the driver's only perceptual cue or imply that every driver follows an identical rule.
The fitted delay also includes vehicle response and signal processing, so it should not be interpreted as a measured perceptual reaction time.

\subsection{Excess Looming Before Takeover}
\label{sec:model:reference}

At each pre-takeover time $t$, the same calibrated law is initialized at the observed state one second earlier, $u=t-1$~s:
$v_H(u)=v_E(u)$, $s_H(u)=s(u)$, and $a_H(u)=a_E(u)$.
Its delay buffer is initialized with the observed history on $[u-\tau,u]$.
We integrate the response to $t$ using lead motion observed over $(u,t]$ and obtain the reference looming $\hat\Lambda_H(t\mid t-1)$.
The recorded and simulated trajectories therefore begin from the same state and encounter the same lead motion over that second.
All inputs are available by the assessment time $t$.

Excess looming is the positive difference between observed looming and this reference:
\begin{equation}
D_\Lambda(t)=\max\!\left\{0,\Lambda_A(t)-\hat\Lambda_H(t\mid t-1)\right\}.
\label{eq:dlam}
\end{equation}
A larger $D_\Lambda$ means that the vehicle is closing on the lead more quickly, relative to its gap, than the fitted human response predicts.
For example, if the model calls for braking during the preceding second but the observed vehicle continues approaching at nearly the same speed, excess looming can increase.
This quantity measures the difference from the response seen in the calibration data; it is not a measurement of what a particular driver expects.

\subsection{Association with Disengagement}
\label{sec:methods:hazard}
\label{sec:model:timing}

We compare the occurrence of disengagement across the closing episodes described in Section~\ref{sec:methods:ccdataset}.
Each episode is followed from closing onset for up to three seconds, ending earlier if ADAS disengages or closing ceases.
The primary analysis requires no driver input in the three seconds before onset.
Inputs after onset remain part of the observed takeover process.
The onset definition and sample construction are given in Appendix~\ref{app:robustness}.

For each interval $(t_k,t_k+0.25]$, the model uses the state at its left edge:
\begin{equation}
\operatorname{logit}h(t_k)=\alpha(d)+f_1(c)+f_2(\Lambda_A)
+\beta_3\dot\Lambda_A+\beta_4D_\Lambda+\mathbf z^\top\boldsymbol\eta.
\label{eq:hazard}
\end{equation}
Here $d$ is elapsed time from onset; $\alpha$, $f_1$, and $f_2$ are restricted splines; and $\mathbf z$ contains ego speed, log source, and lead-quality controls.
The three looming terms represent current urgency, its rate of change, and its excess over the response reference.
We fit a ridge-penalized logistic model on the spline design, standardizing linear terms with training-fold statistics.
Driver inputs are used to define samples and event classes, but are excluded from the predictors.

Training and testing use different drivers, and the GHR reference for each test driver is fitted without that driver's data.
Model selection uses only the training data.
Supporting comparisons and uncertainty estimates are reported in Appendix~\ref{app:robustness}.
\authorcheck{M2}{Confirm the onset-confirmation timestamp, inner-fold isolation of the response reference, spline specification, penalty selection, and covariance calculation after standardization against the analysis code. The supplied tables report these estimates but the fitting code is absent.}

The primary analysis contains 1,887 episodes and 308 disengagements.
Greater excess looming is associated with a higher likelihood of takeover.
For each $0.01~\mathrm{s^{-1}}$ increase in excess looming, the estimated odds of disengagement in the next quarter-second are about 20\% higher after accounting for the other variables.
The association remains positive in the broader sample, in comparisons within the same drive, and when takeover is anchored to pedal input (Table~\ref{tab:lm_association}).

Adding excess looming raises interval-level AUROC from 0.756 to 0.765.
This is a modest improvement in discrimination.
Comparisons with models using recent kinematic history are given in Appendix~\ref{app:robustness}; they do not establish a predictive advantage over a flexible tree model.
The response model is useful because it expresses the recent approach in terms of braking behavior and makes the difference between assisted and human responses easier to interpret.

\section{Discussion}
\label{sec:discussion}
\label{sec:model:limits}

ADAS-TO allows us to follow a takeover from the driver's first intervention to the change in ADAS status and the vehicle response afterward.
The observations distinguish the start of driver intervention from the subsequent change in longitudinal conditions.
This distinction matters when evaluating whether a takeover has relieved an ongoing conflict.

\subsection{From Driver Intervention to Longitudinal Recovery}

In almost two-thirds of takeovers involving braking or steering, the driver starts intervening before ADAS disengages.
A driver may already be pressing a pedal or applying steering while the system still reports that it is engaged.
The first input marks the beginning of the driver's intervention; disengagement marks the system's transition out of assisted control.
Their separation is usually short and is influenced by the system's input and disengagement logic.
Both timestamps are needed to describe the transfer of control.

After disengagement, braking must change the relative motion of the two vehicles.
In brake-involved events, the mean TTC curve reaches its minimum about 1.2~s after disengagement.
The paired-event results also show that lower TTC values frequently occur in the following five seconds.
These observations are consistent with a period in which the driver's response is developing while the closing conflict continues.
The 1.2-s value locates the minimum of a population mean curve and does not measure an individual driver's reaction time.

Continued closing does not by itself require TTC to decrease.
For the same lead vehicle, let $b_{\mathrm{rel}}=a_L-a_E$ denote relative deceleration.
Since $\dot s=-c$ and $\dot c=-b_{\mathrm{rel}}$, while $c>0$,
\begin{equation}
\frac{\mathrm d\,\mathrm{TTC}}{\mathrm dt}
=-1+\frac{s\,b_{\mathrm{rel}}}{c^2}.
\label{eq:ttc_change}
\end{equation}
The gap continues to shrink while the ego vehicle is faster than the lead, but TTC begins to increase once relative deceleration exceeds $c^2/s$.
Thus, TTC can improve before the two speeds become equal.
The observed timing depends on the driver's developing braking response, vehicle dynamics, and lead motion.
It cannot be attributed to a fixed human response delay alone.

\subsection{Implications for Takeover and Collision-Avoidance Systems}

Takeover safety should be evaluated through the subsequent vehicle response.
The first driver input establishes that intervention has begun, while the following motion shows whether it is sufficient to manage the closing conflict.
Measures at disengagement should therefore be considered alongside subsequent minimum TTC, braking demand, and recovery of relative motion.

This distinction is relevant to collision-avoidance support during a transfer to manual control.
Dynamic brake support already supplements driver braking when it is insufficient to avoid an impending collision \citep{nhtsa_driver_assistance}.
The present observations provide a reason to examine that need throughout the developing response after ADAS disengagement.
Such assessment would use the available gap, lead motion, and the braking being applied to project how the approach will develop.
A falling TTC alone does not establish that a collision is unavoidable, because TTC assumes unchanged velocities while braking is changing them.

The minimum of the mean TTC curve identifies a period deserving attention in this dataset; it does not specify a universal waiting period or an AEB trigger.
Deriving an intervention threshold would require event-level prediction of relative motion and validation of the resulting braking strategy.

\subsection{Interpreting the Human-Response Model}

GHR gives a compact description of post-takeover braking and reproduces spacing on held-out drivers about as well as IDM.
Its looming formulation connects the model with evidence that human braking depends on the urgency of the approach \citep{markkula2016farewell,svard2021computational}.
The fitted parameters describe how speed and spacing modify this response.
This combination of trajectory fit and physical interpretation supports its use as a human-response reference.

Greater excess looming is associated with a higher likelihood of disengagement.
In the sampled episodes, drivers are more likely to take over when the approach produces greater looming than the response observed after takeover would predict.
This links the preceding assisted approach with a model calibrated on manual driving.
The reference summarizes behavior across the calibration sample; it does not recover each driver's internal expectation or establish the perceptual cause of intervention.

The reference adds a small amount of discrimination to the spline model.
Other following laws and lead-motion assumptions produce similar gains, and a flexible model of recent kinematics can capture much of the same information.
GHR is useful here because the recent approach can be interpreted through an explicit braking response.
Further evaluation of predicted TTC minima and their timing would test how well its trajectory fit extends to the transition characteristics observed in Section~\ref{sec:stats}.

\subsection{Scope and Limitations}

The dataset includes many drivers and vehicle models, but their contributions are uneven.
Evaluation on held-out drivers tests transfer beyond the drivers used for calibration, while the collected events still reflect the driving conditions represented in this corpus.
Maneuver filtering helps distinguish following responses from planned actions, although manual context labels remain subject to reviewer disagreement.
The safety-critical cases describe the situations found in that reviewed set rather than their prevalence in all assisted driving.

The event time is ADAS disengagement, which may follow a driver input, a cancellation, or a system exit.
Input timing helps interpret these events but does not identify the initiating cause of every exit.
Small differences between input and disengagement timestamps also depend on signal timing and processing.
The closing-episode analysis also uses different onset-selection rules for cases and controls.
Its associations apply to the sampled episodes; absolute takeover probabilities and false alarms per driving hour require continuous driving exposure.

The response model is calibrated after takeover and then used as a reference for the preceding approach.
This assumes that the observed manual responses provide a useful comparison for the earlier state.
TTC and looming also depend on the tracked lead, which can change or become unavailable.
Lead loss may reflect a change in the traffic relationship or a gap in the recorded lead state.
The recovery estimates rely on assumptions about these unobserved outcomes, and the same-lead analysis is conditional on continued tracking.
Reviewing the reasons for lead loss and testing alternative censoring assumptions would strengthen population-level estimates of recovery.
The observations do not establish how the conflict would have developed had ADAS remained active.

\section{Conclusion}
\label{sec:conclusion}

ADAS-TO provides synchronized video and vehicle signals for 16,446 naturalistic ADAS takeovers.
The recordings cover 364 drivers and 99 vehicle models, and manual review identifies 6,206 events for studying takeovers during continued lane keeping or car following.

The observations show that takeover is a process that extends beyond the change in ADAS status.
Drivers often begin braking or steering while the system is still active.
Lower TTC values are frequently observed after takeover, and the mean TTC curve in brake-involved events reaches its minimum after disengagement.
The driver's developing response and the resulting change in relative motion therefore need to be included in takeover safety assessment.
The data also reveal recurring longitudinal contexts, including rapid closing, cut-ins, and discrepancies in the recorded lead representation.

A GHR model fitted to post-takeover trajectories provides a compact description of braking in terms of looming, speed, and spacing, with trajectory performance close to IDM on held-out drivers.
The same model provides a reference for the preceding approach: disengagement is more likely when observed looming exceeds the response predicted from the human-driven trajectories.
These results connect the conditions preceding takeover with the subsequent braking response and motivate continued assessment of longitudinal safety as control passes to the driver.

\section*{Acknowledgements}
This research was supported by the start-up fund of Dr. Hao Zhou at the University of South Florida. The authors thank the autonomous driving community contributors who voluntarily shared driving data and helped make large-scale naturalistic ADAS takeover analysis possible.

\section*{Data availability}

The ADAS-TO dataset and both study-specific subsets are publicly
available on Hugging Face. The full dataset, including
takeover-centered front-view video clips, synchronized vehicle
and perception signals, and event metadata, is available at
\url{https://huggingface.co/datasets/HenryYHW/ADAS-TO}.
The safety-critical subset used for the contextual analysis is
available at
\url{https://huggingface.co/datasets/HenryYHW/ADAS-TO-Critical}.
The longitudinal subset used for car-following response modeling
and the comparison of takeover and continued-ADAS episodes is
available at
\url{https://huggingface.co/datasets/HenryYHW/ADAS-TO_longitudinal}.
All three resources can be downloaded with an access request.

\FloatBarrier
\appendix
\section{Scenario Classification}
\label{app:maneuver_filtered_subset}

Signal-based rules used curvature, steering, turn signals, speed, and stopping cues to identify maneuver candidates.
Reviewers then inspected the full 20-s video with synchronized signals using the criteria in Table~\ref{tab:final_scenario_groups}.
Every clip retained for the primary analysis was reviewed manually; uncertain cases were assigned to Atypical.

\begin{table}[pos=htbp]
\centering
\caption{Scenario labels in the full dataset and their review criteria. Counts describe the labeled corpus before selection of the final analysis set.}
\label{tab:final_scenario_groups}
\small
\renewcommand{\arraystretch}{1.12}
\begin{tabularx}{\linewidth}{@{}p{0.22\linewidth}Xr@{}}
\toprule
Group & Observable criterion & Clips \\
\midrule
Maneuver-Filtered ADAS & Continued lane keeping or following, without a visible maneuver, road-context confound, or unresolved ambiguity. & 6,329 \\
Turning & Intersection turn, departure from the through path, or turn-specific geometry. & 4,192 \\
Lane Changing & Lane crossing, merge, diverge, or other visible lane transition; a turn signal alone is insufficient. & 2,581 \\
Traffic-Control Stopping & Stopping or clear deceleration toward a red light, stop sign, stop line, or similar traffic control. & 3,090 \\
Atypical & Complex geometry, road-surface disturbance, unusual scene, data anomaly, or insufficient evidence. & 254 \\
\midrule
Total & & 16,446 \\
\bottomrule
\end{tabularx}
\end{table}

Gradual bends, minor steering corrections, and low-speed following were retained when no discrete maneuver or traffic-control stop was visible.
Roundabouts, work-zone lane shifts, and road-surface humps were excluded.
The labels describe the visible driving context around the intervention.

\paragraph{Label agreement.}
\label{app:blind_audit}
A blind relabeling audit of 325 clips gave 79.4\% agreement (Cohen's $\kappa=0.53$).
The audit was conducted within the research team, and reviewer identities were not recorded.
The longitudinal analysis uses the final set of 6,206 manually confirmed events.

\begin{figure}[pos=htbp]
    \centering
    \begin{subfigure}[c]{0.54\linewidth}
        \centering
        \includegraphics[width=\linewidth]{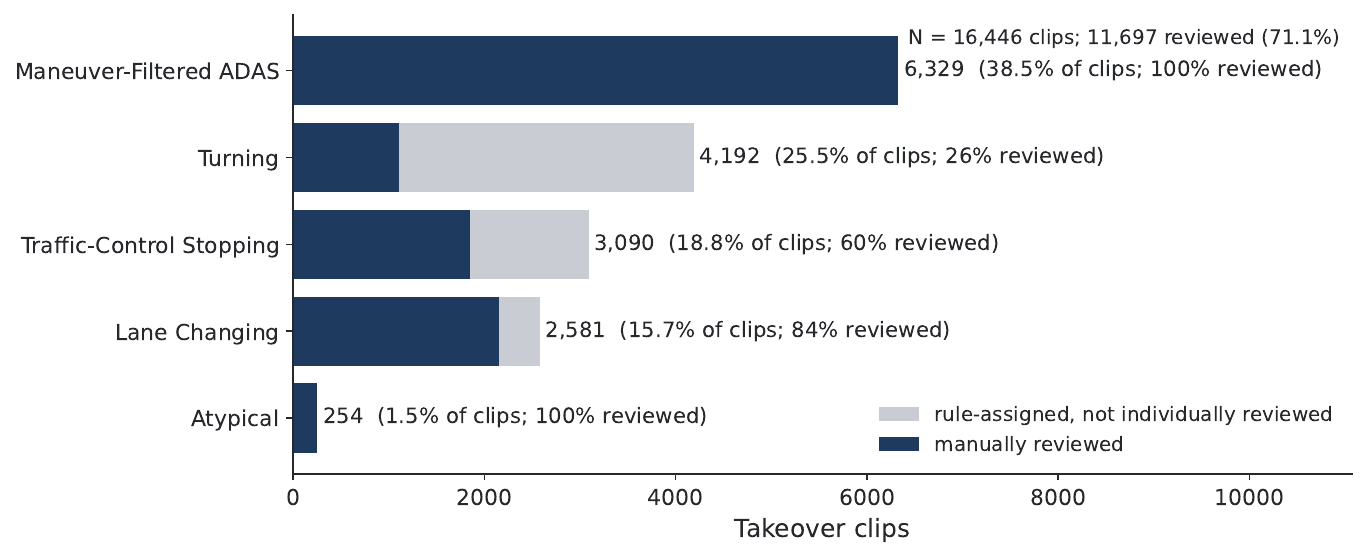}
        \caption{Scenario groups and manual-review coverage.}
        \label{fig:divisions}
    \end{subfigure}\hspace{0.025\linewidth}
    \begin{subfigure}[c]{0.38\linewidth}
        \centering
        \includegraphics[width=\linewidth]{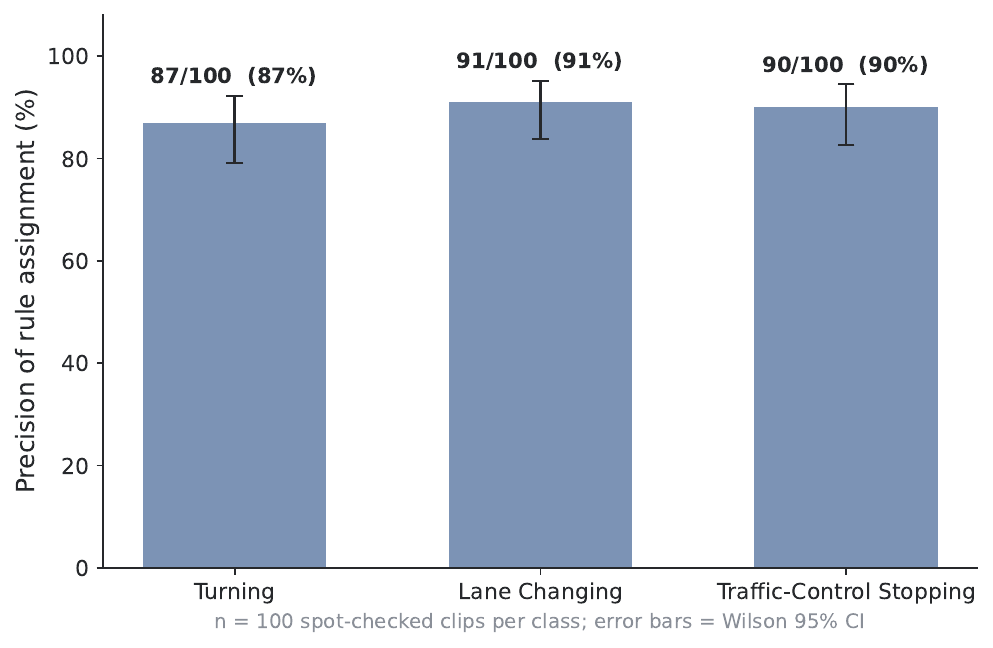}
        \caption{Precision of maneuver-screening rules.}
        \label{fig:rule_audit_precision}
    \end{subfigure}
    \caption{Scenario filtering in ADAS-TO. (a) Scenario groups and manual-review coverage. (b) Precision of the maneuver-screening rules, with Wilson uncertainty intervals.}
    \label{fig:scenario_filtering}
\end{figure}

\section{Selection of Safety-Related Cases}
\label{app:safety_relevant_cases}

The context-review screening population comprises 5,781 takeovers with ego speed above 3~m/s at disengagement.
An event was selected if it met any indicator in Table~\ref{tab:app_selection}; signal availability determines each row's valid sample.
Percentile thresholds use the eligible distribution and select approximately its most extreme 2\%.
Manual context labeling covers 749 reviewed cases.
Both indicator membership and the labels in Table~\ref{tab:all_review_contexts} are non-exclusive. This context-review inventory is reported separately from the revised descriptive cohort; the supplied context counts are retained.

\begin{table}[pos=htbp]
\centering
\caption{Indicators used to select cases for manual review. Windows are relative to disengagement; $P_{98}$ denotes the eligible-sample 98th percentile.}
\label{tab:app_selection}
\small
\setlength{\tabcolsep}{4pt}
\renewcommand{\arraystretch}{1.12}
\begin{tabularx}{\linewidth}{@{}Xp{1.6cm}p{4.2cm}rr@{}}
\toprule
Indicator & Window (s) & Threshold & Valid $n$ & Selected \\
\midrule
Minimum TTC & $[0,5]$ & $<2.0$~s & 3,662 & 142 \\
Maximum 1-s TTC decline & $[0,5]$ & $>P_{98}$ ($\approx20.5$) & 2,655 & 54 \\
Minimum longitudinal acceleration & $[0,3]$ & $<-5.0$~m/s$^2$ & 5,781 & 85 \\
$P_{95}$ longitudinal jerk magnitude & $[0,2]$ & $>P_{98}$ ($\approx9.4$~m/s$^3$) & 5,781 & 116 \\
Maximum planar acceleration, normalized by $g$ & $[0,5]$ & $>0.5$ & 5,526 & 148 \\
Minimum TLC & $[0,5]$ & $<1.0$~s & 3,820 & 151 \\
Maximum absolute lane offset & $[-3,3]$ & $>0.85$~m & 4,014 & 253 \\
$P_{95}$ yaw-acceleration magnitude & $[0,2]$ & $>P_{98}$ ($\approx0.16$~rad/s$^2$) & 5,521 & 111 \\
Maximum steering-wheel rate & $[0,3]$ & $>P_{98}$ ($\approx81^\circ$/s) & 5,781 & 116 \\
\bottomrule
\end{tabularx}
\end{table}

Lead-based measures require a valid logged lead; lane measures require lane confidence above 0.5.
Acceleration and angular-rate measures use the calibrated vehicle-body frame (Appendix~\ref{app:measures}).
Reviewers labeled visible traffic and road contexts from the video and synchronized signals.
A lead-representation discrepancy was coded within rapid-closing events when the visible lead and its logged representation were inconsistent.

\begin{table*}[pos=htbp]
\centering
\caption{Manual context labels in the manually reviewed safety-critical set ($n=749$). Labels are multi-label; totals may exceed one hundred percent.}
\label{tab:all_review_contexts}
\small
\renewcommand{\arraystretch}{1.10}
\begin{tabular}{p{0.31\textwidth}rr@{\hspace{1.8em}}p{0.31\textwidth}rr}
\hline
\textbf{Longitudinal context} & \textbf{$n$} & \textbf{Percent} &
\textbf{Lateral context} & \textbf{$n$} & \textbf{Percent} \\
\hline
Rapid closing towards lead vehicle & 323 & 43.1 &
Sharp curvature & 122 & 16.3 \\
\hspace{1em}Lead-representation discrepancy$^\dagger$ & 97 & 13.0 &
Changing road geometry & 69 & 9.2 \\
Sudden cut-in by surrounding vehicles & 82 & 10.9 &
Low lane-line visibility & 58 & 7.7 \\
Danger sign or warning context & 16 & 2.1 &
Abnormal lateral path swaying & 20 & 2.7 \\
& & &
Nearby large or close-proximity vehicle & 107 & 14.3 \\
& & &
Oncoming traffic & 29 & 3.9 \\
\hline
\multicolumn{6}{l}{\textit{Other observed context}} \\
\hline
Vertical bump or road disturbance & 14 & 1.9 &
Other & 30 & 4.0 \\
\hline
\end{tabular}
\vspace{2pt}
\begin{minipage}{\textwidth}
\footnotesize
\textit{Note.}
$^\dagger$ An apparent lead-representation discrepancy denotes a case in which a visually relevant in-path lead vehicle was absent from, delayed in, or inconsistent with the available logged lead representation.
This label is coded within rapid-closing events and describes the available video and log evidence.
Changing road geometry combines merge/diverge and lane-line-change contexts.
\end{minipage}
\end{table*}

\section{Driver Intervention Behavior Patterns}
\label{app:patterns}
\label{sec:driver_input_evolution}

This appendix summarizes the average driver-input pattern around Maneuver-Filtered ADAS interventions.
This analysis complements the preceding longitudinal and lateral results by showing how the main control channels evolve together around the disengagement edge.

\begin{figure*}[pos=htbp]
\centering
\includegraphics[width=0.75\textwidth]{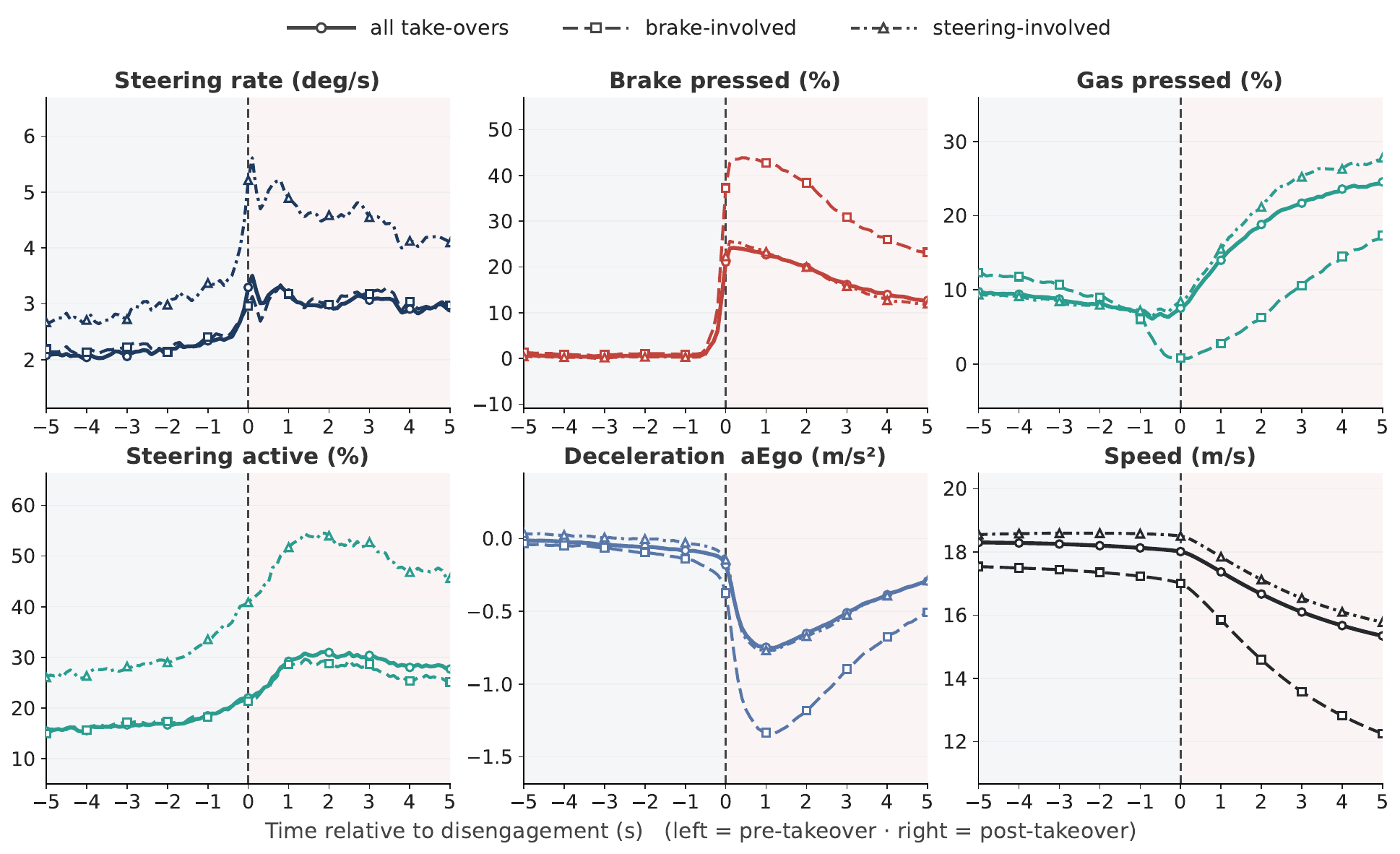}
\caption{
Population-level driver input and vehicle response around Maneuver-Filtered ADAS interventions.
Solid lines show all takeovers, dashed lines show brake-involved takeovers, and dash-dotted lines show steering-involved takeovers.
Binary input channels are reported as the fraction of active events per second; continuous variables are reported as per-second means.
The vertical dashed line marks the disengagement edge at $t=0$.
}
\label{fig:driver_input_timecourse}
\end{figure*}

Fig.~\ref{fig:driver_input_timecourse} shows three distinct input patterns.
First, braking is pulse-like: brake activity and deceleration rise sharply around disengagement, peak shortly after intervention, and then decline.
Second, steering is more sustained: steering activity also peaks after disengagement, but remains elevated for several seconds, indicating continued lateral correction.
Third, gas input increases gradually after intervention, consistent with speed recovery or continued accelerator override in part of the events.

The subset curves show similar differences.
Brake-involved interventions show a sharper braking pulse, stronger deceleration, and larger speed reduction, whereas steering-involved interventions show higher steering activity and steering rate throughout the transition window.
Thus, the average driver response is not completed at the disengagement instant.
In these event averages, braking is brief, steering remains active for longer, and accelerator input becomes more common later in the observation window.


\section{Measures of Driver Response, Safety, and Vehicle Motion}
\label{app:measures}

\begin{table}[pos=htbp]
\centering
\caption{Temporal alignment and analysis windows.}
\label{tab:analysis_windows}
\small
\renewcommand{\arraystretch}{1.15}
\begin{tabularx}{\linewidth}{@{}p{3.0cm}p{2.5cm}X@{}}
\toprule
\textbf{Component} & \textbf{Time range} & \textbf{Primary purpose} \\
\midrule
Context window
& $[-10,+10]$~s
& Complete event-centered clip released in ADAS-TO. \\

Main analysis window
& $[-5,+5]$~s
& Common interval used for event-level comparisons. \\

Pre-take-over window
& $[-5,0)$~s
& Characterizes traffic, lane, and vehicle conditions before ADAS disengagement. \\

Input-to-disengagement interval
& $T_{\mathrm{input}}$ relative to $0$
& Determines whether brake or steering input occurs before, at, or after disengagement. \\

Early post-take-over window
& $[0,+2]$~s
& Captures the immediate braking, steering, and vehicle response. \\

Late post-take-over window
& $(+2,+5]$~s
& Characterizes continued safety-margin changes and vehicle-motion recovery. \\
\bottomrule
\end{tabularx}
\end{table}

\paragraph{Driver-input measures.}
Braking and acceleration come from pedal states and analog pedal signals; steering counts as active when driver input is detected with an absolute wheel rate above $2^\circ/\mathrm{s}$; channels are non-exclusive.
Three input descriptions are used: the onset of the first brake or steering action ($T_{\mathrm{input}}$, for timing; the contiguous active block is traced back to its start, and gas is excluded), the channel active within $[-0.3,+0.1]$~s of the disengagement edge, and the channels active within $[0,+2]$~s after it.

\paragraph{Longitudinal safety measures.}
For clips with a valid radar lead (\texttt{leadOne.status} $>0.5$ and $d_{\mathrm{rel}}>0$), we measure lead distance, closing speed, and TTC (computed only while closing and kept when $0.1<\mathrm{TTC}<90$~s), at the disengagement edge and as extrema over the symmetric windows $[-5,0)$~s and $[0,+5]$~s (minimum TTC, peak closing speed, minimum distance).
Post-takeover deceleration peaks above 9.8~m/s$^2$ are excluded as a signal-quality screen.
\authorcheck{S1}{Complete the pending review of the 10 high-deceleration cases, document the sensor-validation rule, and report sensitivity to retaining verified observations.}
For brake-involved takeovers, observed braking is compared with the demand implied by the instantaneous closing geometry:

\begin{equation}
a_{\mathrm{req}}
=
\frac{\Delta v^2}{2d},
\end{equation}

where $\Delta v>0$ is relative speed and $d$ is lead distance; the braking margin is $a_{\mathrm{actual}} - a_{\mathrm{req}}$ with $a_{\mathrm{actual}}$ the post-takeover peak deceleration.
This instantaneous kinematic reference assumes unchanged lead speed and omits response delay and road grade.

\paragraph{Persistence and resolution.}
The paired persistence analysis compares the minimum valid TTC in $[-5,0)$~s with the minimum in $[0,5]$~s for events observed in both windows.
Separate summaries restrict the sample to events closing at the disengagement edge and to events with the same lead for the next 5~s.
This comparison identifies events with a lower post-takeover minimum and does not require TTC to decrease monotonically throughout the window.
The approximately 1.2-s timing in Fig.~\ref{fig:longitudinal_transit} refers to the minimum of the event-aligned mean TTC curve for brake-involved events.
It is distinct from a distribution of individual-event minimum times and from the resolution measure below.
Resolution is defined as the first sustained second during which TTC returns to its pre-edge level or closing speed is non-positive.
The criterion describes recovery relative to the earlier longitudinal state or cessation of closing; it does not require a specified absolute safety margin.
Follow-up is censored at ADAS re-engagement, lead loss, or 5~s.
Kaplan--Meier estimates describe resolution under the assumption that censoring is non-informative; lead loss may violate that assumption.
\authorcheck{S2}{Specify how the pre-edge TTC reference is calculated and whether the recovery timestamp is the start of the sustained one-second interval or its confirmation time. Verify these choices against the code before changing the definition or reported estimates. Review lead-loss reasons and assess sensitivity to informative censoring. These checks have not been performed in this wording revision.}

\paragraph{Lateral safety measures.}
Lateral safety is evaluated using time-to-line-crossing (TLC) when valid lane geometry is available.
TLC is calculated during outward lateral motion as

\begin{equation}
\mathrm{TLC}
=
\frac{
w/2-\left|d_{\mathrm{lane}}\right|
}{
\left|\dot{d}_{\mathrm{lane}}\right|
},
\end{equation}

where $w$ is lane width and $d_{\mathrm{lane}}$ the lateral position relative to lane center.
TLC and the maximum absolute lane deviation are computed at lane-line confidence above 0.5; lane deviation describes the lateral trajectory.

\paragraph{Vehicle-response measures used in context screening.}
IMU signals are transformed to the vehicle-body frame with the recorded calibration.
Screening measures include longitudinal jerk $j_x=\mathrm{d}a_x/\mathrm{d}t$, yaw acceleration $\dot r=\mathrm{d}r/\mathrm{d}t$, and planar acceleration normalized by gravity,
\begin{equation}
a_{\mathrm{comb}}=\frac{\sqrt{a_x^2+a_y^2}}{g}.
\end{equation}
Derivative-based measures use 0.2-s smoothing.
Indicator windows and thresholds are listed in Table~\ref{tab:app_selection}.
All lead- and lane-based measures require their stated signal-validity conditions; invalid observations are treated as missing.

\section{Supporting Model Analyses}
\label{app:robustness}

This appendix records sample construction, predictor comparisons, and sensitivity analyses supporting Section~\ref{sec:model}.
All splits and uncertainty calculations use drivers as the grouping unit.
Bracketed uncertainty ranges in Tables~\ref{tab:post_summary} and~\ref{tab:lm_association}--\ref{tab:family_compact} represent 95\% confidence intervals.
Association estimates use driver-clustered standard errors; predictive comparisons resample drivers.

\subsection{Samples and Event Definitions}

The initial screen requires closing speed of at least 0.5~m/s for one second in $[-3,0]$~s and at least 90\% lead coverage in $[-3,+1]$~s around disengagement.
Closing onset is confirmed after four consecutive 0.25-s bins at or above the speed threshold, with at least one second of preceding history.
Cases use the episode still ongoing at disengagement within the preceding 9.8~s.
Controls use the earliest eligible onset in a 10-s engaged, input-free segment from the same drive, with no disengagement within 60~s.
Up to two controls are sampled per case, targeting a starting-speed difference within 5~m/s; this restriction is relaxed for 37.4\% of sampled controls.
Both groups are followed for at most three seconds, ending earlier at disengagement or a closing lapse.

Starting with 2,293 screened closing cases, revised scenario labels remove five.
Further exclusions are an unobserved onset (666), inadequate lead coverage (218), and no closing episode ongoing at disengagement (113), leaving 1,291 candidates.
The control search yields 5,200 eligible onset windows after the drive and disengagement guards; 1,272 are sampled.
Complete-feature filtering and three-second follow-up produce the samples in Table~\ref{tab:cohort_final}.
A candidate whose disengagement lies after the horizon contributes censored exposure, not an event.

\begin{table}[pos=htbp]
\centering
\caption{Final modeling samples after complete-feature requirements. Events are disengagements observed within three seconds of onset.}
\label{tab:cohort_final}
\small
\begin{tabular}{lrrrrr}
\toprule
Sample & Case windows & Controls & Intervals & Events & Drivers \\
\midrule
Full & 1,215 & 1,268 & 21,802 & 631 & 205 \\
Input-clean & 619 & 1,268 & 17,013 & 308 & 182 \\
\bottomrule
\end{tabular}
\end{table}

The input-clean rule applies to the three seconds before onset and permits input afterward.
It differs from an earlier rule excluding longitudinal input throughout the ten seconds before disengagement.
The matched analysis retains cases with same-drive controls.
Median onset speed differs between case candidates and controls (17.2 versus 22.5~m/s); ego speed remains a covariate.
Cases use the onset of the closing episode ongoing at the event, whereas controls use the earliest eligible onset in their segment.
These rules remain relevant when interpreting conditional associations.

The 308 primary disengagements include 209 brake-led events, 15 accelerator-led events, 18 steering-led events without pedal input, 54 with input only before the two-second interval preceding disengagement, and 12 with no recorded input in $[-10,+2]$~s.
These classes describe the available signals rather than determine the initiating cause of every exit.
The first-pedal-input sensitivity changes the anchor and yields 316 events.

\subsection{Association Across Specifications}

Table~\ref{tab:lm_association} reports the primary, full, and matched associations, together with input-anchor and contributor-exclusion analyses.
Each specification includes the covariates in Eq.~\eqref{eq:hazard}.
In the primary sample, $D_\Lambda$ has median $0.0036$ and 90th percentile $0.0146~\mathrm{s^{-1}}$, and is positive in 79\% of intervals.
Its coefficient is 18.5 [9.9, 27.1], giving an odds ratio of 1.203 per $0.01~\mathrm{s^{-1}}$.
A driver bootstrap gives an odds-ratio interval of 1.125--1.352.
These odds ratios concern a 0.25-s interval conditional on its covariates.

\begin{table}[pos=htbp]
\centering
\caption{Association between excess looming and disengagement across sample definitions. Odds ratios are per $0.01~\mathrm{s^{-1}}$ increase in $D_\Lambda$; intervals are driver-clustered uncertainty ranges.}
\label{tab:lm_association}
\small
\renewcommand{\arraystretch}{1.12}
\begin{tabularx}{\linewidth}{@{}Xrc@{}}
\toprule
Sample or event definition & Events & Odds ratio [range] \\
\midrule
Primary input-clean sample & 308 & 1.203 [1.105, 1.311] \\
Full sample & 631 & 1.123 [1.059, 1.190] \\
Same-drive matched sample & 323 & 1.125 [1.058, 1.196] \\
Input-clean, event at first pedal input & 316 & 1.200 [1.102, 1.308] \\
Input-clean, five most prolific drivers excluded & 233 & 1.240 [1.123, 1.370] \\
\bottomrule
\end{tabularx}
\end{table}

\subsection{Kinematic History and Predictive Comparisons}

R0 contains Eq.~\eqref{eq:hazard} without $D_\Lambda$.
R0L adds one second of lagged kinematics.
R4 and R4L add $D_\Lambda$ to R0 and R0L.
Table~\ref{tab:prediction_compact} reports held-out interval-level discrimination and log-loss; Table~\ref{tab:contrasts_compact} reports paired differences with 2,000 driver bootstrap resamples.
Duration-stratified concordance compares pairs at the same elapsed duration.

\begin{table}[pos=htbp]
\centering
\caption{Selected predictor comparisons under strict outer-fold isolation. NLL is interval negative log-likelihood; lower values are better.}
\label{tab:prediction_compact}
\small
\setlength{\tabcolsep}{4pt}
\begin{tabular}{llccc}
\toprule
Sample & Model & AUROC [range] & \shortstack{Duration-stratified\\concordance} & NLL ($10^{-3}$) \\
\midrule
Full & R0 & 0.724 [0.697, 0.753] & 0.726 & 121.29 \\
 & R0L & 0.718 [0.692, 0.746] & 0.719 & 121.87 \\
 & R4 & 0.729 [0.701, 0.758] & 0.731 & 120.94 \\
 & R4L & 0.726 [0.699, 0.754] & 0.727 & 121.24 \\
\midrule
Input-clean & R0 & 0.756 [0.722, 0.789] & 0.750 & 82.29 \\
 & R0L & 0.745 [0.704, 0.778] & 0.737 & 82.81 \\
 & R4 & 0.765 [0.731, 0.798] & 0.759 & 81.57 \\
 & R4L & 0.759 [0.727, 0.797] & 0.752 & 81.74 \\
\bottomrule
\end{tabular}
\end{table}

\begin{table}[pos=htbp]
\centering
\caption{Paired changes in AUROC and NLL. Changes are the second model minus the first; each contrast uses the same evaluation sample.}
\label{tab:contrasts_compact}
\small
\setlength{\tabcolsep}{4pt}
\begin{tabularx}{\linewidth}{@{}lXcc@{}}
\toprule
Sample & Comparison & $\Delta$AUROC [range] & $\Delta$NLL ($10^{-3}$) [range] \\
\midrule
Full & R0 to R4 & +0.005 [0.002, 0.009] & $-0.35$ [$-1.09$, 0.44] \\
 & R0L to R4L & +0.008 [0.004, 0.014] & $-0.62$ [$-1.35$, 0.18] \\
 & XGB R0L to GAM R4 & $-0.009$ [$-0.021$, 0.005] & +0.22 [$-1.37$, 1.72] \\
\midrule
Input-clean & R0 to R4 & +0.009 [0.002, 0.017] & $-0.72$ [$-1.39$, $-0.10$] \\
 & R0L to R4L & +0.014 [0.005, 0.024] & $-1.07$ [$-1.93$, $-0.11$] \\
 & XGB R0L to GAM R4 & +0.002 [$-0.013$, 0.016] & $-0.43$ [$-1.60$, 0.80] \\
\bottomrule
\end{tabularx}
\end{table}

Excess looming improves the spline model with and without the selected lagged variables.
Neither the AUROC nor the NLL comparison establishes an advantage over the gradient-boosted tree model supplied with lagged kinematics.
The comparison supports an interpretable summary of history, without establishing information unavailable to a more flexible history-based predictor.

\subsection{Response-Law and Lead-Input Comparisons}

We construct alternative references with IDM and Gipps, and replace observed lead motion with extrapolations from the start of the one-second reference window.
The lead alternatives are constant velocity, constant acceleration, exponentially decaying acceleration, and free-flow IDM recovery toward a recent speed.
Observed motion holds the traffic input common to the recorded and simulated ego responses; predicted inputs also introduce the difference between predicted and realized lead behavior.

The AUROC gains in Table~\ref{tab:family_compact} occur with several laws and lead inputs.
They do not identify a uniquely preferred law or a particular mental forecast used by the driver.
GHR is retained for its compact form and explicit relation to looming.

\begin{table}[pos=htbp]
\centering
\caption{Selected reference constructions: change in interval AUROC over R0 [range] under the onset-aligned evaluation protocol.}
\label{tab:family_compact}
\small
\setlength{\tabcolsep}{4pt}
\begin{tabularx}{\linewidth}{@{}lXcc@{}}
\toprule
Law & Lead input & Input-clean & Full \\
\midrule
GHR & Observed motion & +0.009 [0.002, 0.017] & +0.005 [0.002, 0.009] \\
IDM & Observed motion & +0.010 [0.004, 0.017] & +0.008 [0.004, 0.014] \\
Gipps & Observed motion & +0.009 [0.003, 0.015] & +0.005 [0.001, 0.011] \\
\midrule
GHR & Constant velocity & +0.014 [0.008, 0.022] & +0.010 [0.006, 0.015] \\
GHR & Constant acceleration & +0.009 [0.000, 0.019] & +0.004 [0.001, 0.007] \\
GHR & Decaying acceleration & +0.013 [0.003, 0.025] & +0.006 [0.002, 0.010] \\
GHR & Free-flow IDM recovery & +0.007 [0.002, 0.013] & +0.006 [0.002, 0.010] \\
\bottomrule
\end{tabularx}
\end{table}

\subsection{Alarm Operating Points and Sensitivity}

Alarm results are secondary to the behavioral analysis.
Thresholds are selected from training-fold controls and evaluated on held-out groups.
Interval thresholds and thresholds on the window maximum have different false-alarm denominators.
Table~\ref{tab:alarm_compact} reports R4 with thresholds on window maxima.
Only the 308 disengagements observed within the horizon are eligible events.
Censored case windows are not added to either the hit or false-alarm denominator.
The early-detection share includes misses; lead-time summaries include triggered events only.

\begin{table}[pos=htbp]
\centering
\caption{R4 alarm performance in the input-clean sample, with thresholds on training-control window maxima. Parentheses give evaluation counts.}
\label{tab:alarm_compact}
\small
\setlength{\tabcolsep}{4pt}
\begin{tabularx}{\linewidth}{@{}cccccX@{}}
\toprule
Target FA & Window FA & Events detected & At least 1~s early & \shortstack{Median lead\\(s)} & IQR (s) \\
\midrule
2\% & 2.3\% (29/1,268) & 33.1\% (102/308) & 19.8\% (61/308) & 1.00 & [0.75, 1.75] \\
5\% & 5.3\% (67/1,268) & 48.0\% (148/308) & 29.2\% (90/308) & 1.25 & [0.69, 1.75] \\
10\% & 11.6\% (147/1,268) & 57.8\% (178/308) & 38.3\% (118/308) & 1.25 & [0.75, 1.75] \\
\bottomrule
\end{tabularx}
\end{table}

At a 5\% target based instead on interval hazards, the observed interval false-alarm rate is 5.3\%, but window false-alarm rate is 10.0\%.
The associated detection share is 57.1\% (176/308).
These are different operating points; neither is a false-alarm rate per driving hour.

With a 2-s or 4-s horizon, input-clean odds ratios are 1.241 [1.121, 1.374] and 1.222 [1.121, 1.333], respectively.
The AUROC increment at 4~s is 0.003 [$-0.005$, 0.012], so a positive association need not yield a detectable ranking gain in every specification.
Controller-specific refits give increments of 0.010 [0.002, 0.021] for configured openpilot longitudinal control and 0.002 [$-0.011$, 0.014] for stock ACC in the input-clean sample.
These subgroup estimates do not establish a difference between controllers.

\subsection{Recovery Follow-Up}
\label{app:recovery_followup}

Table~\ref{tab:post_summary} uses the paired definitions in Appendix~\ref{app:measures}.
Among the 2,031 events closing at disengagement, resolution is observed in 1,021.
The other observations are censored at lead loss (772), re-engagement (34), or five seconds (204).
In the 994-event same-lead subset, 764 resolutions are observed, with 26 re-engagements and 204 observations reaching the time limit.
Agreement in TTC persistence across these groups is informative, but complete lead tracking is itself a selection condition.

The Kaplan--Meier estimates use observed resolution times and the follow-up available before censoring.
They assume that events censored at a given time would have the same subsequent resolution distribution as those still observed at that time.
This assumption may fail if lead loss is associated with the severity or development of the conflict.
A lead leaving the ego vehicle's path and an interruption in the recorded lead state can both end tracking, but have different implications for whether the original conflict has ended.
The current estimates censor lead loss without distinguishing these causes.
ADAS re-engagement also ends the period of manual control being studied.
The same-lead estimates describe a selected group with continued lead observation; their similarity to the wider sample does not remove uncertainty about unobserved outcomes.
The driver-clustered ranges reflect sampling uncertainty and do not account for bias from informative censoring.
Review of lead-loss causes and sensitivity to alternative assumptions about the unobserved outcomes would be needed to strengthen the interpretation of recovery for the full sample.

\section{Vehicle Models}
\label{app:vehicle_models}

The dataset covers 99 vehicle model families from 23 makes.
Model years, generations, trim levels, powertrains, and ADAS configurations are combined within each named family.
Thus, for example, the hybrid and plug-in versions of the RAV4 count as one model, while Ioniq, Ioniq 5, and Ioniq 6 remain separate.
F-150 includes Lightning, and the Rivian entries are grouped under the R1 family because some records use the broader R1 name.
Unknown entries are not counted as a vehicle model.

\begin{table}[pos=htbp]
\centering
\caption{Vehicle models after consolidating variants.}
\label{tab:vehicle_models}
\small
\renewcommand{\arraystretch}{1.1}
\begin{tabularx}{\linewidth}{@{}lrX@{}}
\toprule
Make & Models & Model names \\
\midrule
Acura & 4 & ADX, Integra, MDX, TLX \\
Alfa Romeo & 1 & Stelvio \\
Audi & 2 & A4, Q5 \\
BYD & 1 & Atto 3 \\
Chevrolet & 5 & Bolt EUV, Equinox, Silverado, Trailblazer, Volt \\
Ford & 7 & Bronco Sport, Escape, Explorer, F-150, Maverick, Mustang Mach-E, Ranger \\
Genesis & 5 & G70, G80, GV60, GV70, GV80 \\
Honda & 10 & Accord, CR-V, City, Civic, Clarity, Fit, HR-V, Odyssey, Pilot, Ridgeline \\
Hyundai & 9 & Elantra, Ioniq, Ioniq 5, Ioniq 6, Kona, Palisade, Santa Fe, Sonata, Tucson \\
Jeep & 2 & Cherokee, Grand Cherokee \\
Kia & 8 & Carnival, Ceed, EV6, EV9, K8, Niro, Sportage, Stinger \\
Lexus & 4 & ES, LC, RC, RX \\
Mazda & 2 & CX-5, CX-9 \\
Nissan & 2 & Leaf, Rogue \\
Porsche & 1 & Macan \\
Ram & 2 & 1500, Heavy Duty \\
Rivian & 1 & R1 family \\
Skoda & 2 & Kodiaq, Superb \\
Subaru & 6 & Ascent, Crosstrek, Forester, Impreza, Legacy, Outback \\
Tesla & 4 & Model 3, Model S, Model X, Model Y \\
Toyota & 7 & Alphard, Camry, Corolla, Highlander, Prius, RAV4, Yaris \\
Volkswagen & 12 & Atlas, Caddy, Golf, ID.4, Jetta, Passat, Sharan, Taos, Tiguan, Touran, Transporter, up! \\
Volvo & 2 & V40, V60 \\
\midrule
Total & 99 & 23 makes \\
\bottomrule
\end{tabularx}
\end{table}

\bibliographystyle{cas-model2-names}
\bibliography{references}

\end{document}